\documentstyle[preprint,aps]{revtex}
\draft
\newcommand{\bold}[1]{\mbox{\boldmath $#1$}}    
\begin{document}

\title{Formation of superdense hadronic matter in
high energy heavy-ion collisions}
\bigskip
\author{Bao-An Li and Che Ming Ko}
\address{Cyclotron Institute and Physics Department,\\
Texas A\&M University, College Station, TX 77843}
\maketitle

\begin{abstract}
We present the detail of a newly developed relativistic
transport model (ART 1.0) for high energy heavy-ion collisions.
Using this model, we first study the
general collision dynamics between heavy ions
at the AGS energies. We then show that in central collisions
there exists a large volume of sufficiently long-lived superdense hadronic
matter whose local baryon and energy densities exceed the critical
densities for the hadronic matter to quark-gluon plasma transition.
The size and lifetime of this matter are found
to depend strongly on the equation of state.
We also investigate the degree and time
scale of thermalization as well as the radial flow during
the expansion of the superdense hadronic matter.
The flow velocity profile and the
temperature of the hadronic matter at
freeze-out are extracted. The transverse momentum and
rapidity distributions of protons, pions
and kaons calculated with and without
the mean field are compared with each other
and also with the preliminary data from the E866/E802
collaboration to search for experimental observables
that are sensitive to the equation of state.
It is found that these inclusive, single
particle observables depend weakly on the equation of state.
The difference between results
obtained with and without the nuclear mean field is only about 20\%.
The baryon transverse collective flow in the
reaction plane is also analyzed.
It is shown that both the flow parameter and the
strength of the ``bounce-off'' effect are very sensitive to the equation
of state. In particular, a soft equation of state with a compressibility
of 200 MeV results in an increase of the flow parameter
by a factor of 2.5 compared to the cascade case without the mean field.
This large effect makes it possible to distinguish the predictions
from different theoretical models and to detect the signatures of
the quark-gluon plasma which is expected to
significantly soften the equation of state.\\
{\bf PACS number(s): 25.75.+r}
\end{abstract}
\newpage
\section{Motivation}
The main purpose of relativistic heavy-ion collisions
at future RHIC and LHC energies is to create in the laboratory
a quark-gluon plasma (QGP) and to study its properties.
On the other hand, for heavy-ion collisions at Brookhaven's AGS
the emphasis has been in studying the properties of hot
dense hadronic matter and the collision dynamics.
However, the possibility of a phase transition to the QGP
in central heavy-ion collisions at AGS energies
has also been mentioned recently.

At AGS, a vast body of data have already been collected and analysed
during past few years\cite{qmatter}. Comparisons of these data
with the predictions from theoretical models, such as the RQMD \cite{rqmd},
the ARC\cite{arc}, and the QGSM\cite{qgsm},
have revealed much interesting physics. In particular, a picture of
nearly complete stopping of baryons in central heavy-ion collisions
has emerged from these studies. Furthermore,
it has been shown that baryon and energy densities up to 10 times that
in normal nuclei have been reached in these collisions, leading thus to
the suggestion that in central collisions at AGS energies
the QGP may have already been formed \cite{kapusta}.
It has been further shown, based on a relativistic nucleation theory,
that perhaps one in every $10^2$ or $10^3$ events undergoes the phase
transition. However, the above suggestion and many other exotic
phenomena depend crucially on the maximum energy and baryon densities
reached in the collision. In this paper, we will carry out a detailed study
on the formation of a superdense hadronic matter in heavy ion collisions
at AGS energies.

With the available Au beam at $p_{beam}/A$=11.6 GeV/c at AGS,
more systematic study of heavy-ion collisions
are being carried out by several collaboration (e.g.\ \cite{e866n,e877n}).
In addition, systematic and exclusive measurements of the
energy and mass dependence of particle production, correlations
and collective flow effects from Bevalac energies up to
AGS energies will soon be carried out by the EOS collaboration\cite{eosn}.
Besides studying possible new physical phenomena, these
experiments will provide a broader base of data so that
more stringent tests of theoretical models can be made.
Also, comparisons of the experimental data with reliable model
predictions will allow us to learn about
the properties of hot dense hadronic matter formed in
these collisions and help identify new physical phenomena.
Stimulated by the success of the RQMD and the ARC in describing
many available experimental data at AGS energies,
we have recently developed a relativistic transport (ART) model
for heavy-ion collisions at these energies.
A brief report of the model has been given in refs.\cite{li95,keywest},
here we will present the detail of the model (ART 1.0)
and study several important aspects and issues of heavy-ion collisions
at AGS energies.

More specifically, we discuss in section II the detail of
the model ART 1.0 and its inputs.
In section III, we first study the general heavy ion
collision dynamics at AGS energies.
We then discuss the formation of the superdense
hadronic matter and its properties
as well as its dependence on the equation of state.
The degree of thermal equilibrium in the superdense hadronic
matter and the time scale involved
are also discussed. In addition, main features of
the radial flow during the expansion phase, the flow velocity
profile and temperature of the hadronic matter at freeze-out
are also investigated. The model is then used to study the
transverse momentum and rapidity distributions of protons, pions
and kaons with and without including the
mean field in order to identify observables that are sensitive to
the equation of state. These results
are then compared with each other and also with the preliminary data
from the E866/E802 collaboration.
Finally, we carry out a detailed analysis of the baryon transverse flow
in the reaction plane and study its dependence on the equation of state.
The summary and outlook are given in section IV.
Those interested only in the results may turn directly to section III.

\section{A relativistic transport model: ART 1.0}
Based on the well-known Boltzmann-Uehling-Uhlenbeck (BUU) model
(e.g. \cite{bertsch,li91a,li91b}) for
intermediate energy heavy-ion collisions, we have recently
developed a relativistic transport model for heavy ion collisions at
AGS energies \cite{li95,keywest} by including more baryon
and meson resonances and their interactions.
The BUU model has been very successful in studying
heavy-ion collisions at beam energies lower than about 3 GeV/nucleon.
We refer the reader to the review by Bertsch and Das Gupta for more details
of the BUU model\cite{bertsch}.

In the ART model, we have kept the same philosophy and methods
as in the BUU model, but added some new physics and
numerical techniques in order to simulate
heavy-ion collisions at higher energies.
More specifically, we have included in ART 1.0 the following
baryons: $N,~\Delta(1232),~N^{*}(1440),~N^{*}(1535),~\Lambda,~\Sigma$,
and mesons: $\pi,~\rho,~\omega,~\eta,~K$,
as well as their explicit isospin degrees
of freedom. Although antiparticles and heavier mesons
have so far not been included, their production can be studied
perturbatively in the present version of the model.
We plan to include explicitly these particles in an upgraded version
of the model. Both elastic and inelastic collisions among most of these
particles are included as best as we can by using as inputs
the experimental data from hadron-hadron collisions.
In accordance with this philosophy, almost all
parameterized cross sections and angular distributions
that have been used in the BUU model are replaced by empirical expressions
based on the double-logarithmic interpolations of the
experimental data\cite{data1}. However, more than 200 reaction channels
are listed in the CERN data book\cite{data1} for nucleon-nucleon and
pion-nucleon collisions.
We certainly have not fully incorporated all these channels.
Instead, most inelastic hadron-hadron collisions
are modeled through the formation of resonances.
The advantage of this approximation is that the finite lifetime of
these resonances takes into account partially the effects of
the finite formation time for produced secondaries.
In the following, we present details of treating
various hadron-hadron interactions.

\subsection{Inelastic baryon-baryon interactions}
First of all, we have included in the model the following reactions,
\begin{eqnarray}
&&NN\leftrightarrow N\Delta,~NN^{*}(1440),~NN^{*}(1535),\\
&&NN\leftrightarrow \Delta\Delta,~\Delta N^{*}(1440),\\
&&NN\rightarrow NN\rho,~NN\omega,~\Delta\Delta\pi,\\
&&NN\rightarrow \Delta\Delta\rho,\\
&&N\Delta\leftrightarrow NN^{*}(1440),~NN^{*}(1535),\\
&&\Delta\Delta\leftrightarrow NN^{*}(1440),~NN^{*}(1535),\\
&&\Delta N^{*}(1440)\leftrightarrow NN^{*}(1535),
\end{eqnarray}
and those producing kaons as shown later in Eqs.\ (\ref{kaon})-(\ref{kaon2}).
The decomposition of the total inelastic nucleon-nucleon cross section into
different channels shown in the above certainly
involves some uncertainties and assumptions.
Moreover, most of the available data are only for the $pp$ reaction.
For the $np$ reaction we use as much data as possible.
Otherwise, we just assume that
the cross sections are the same as in the $pp$ reaction.
In the following, we discuss the methods we have introduced for
determining these cross sections.

As in refs.\ \cite{li91a,wolf90,dani91}, we use the
parameterization introduced by VerWest et al.\cite{verwest}
for the cross sections of producing a single $\Delta$ or $N^{*}(1440)$
resonance in processes shown in Eq. (1).
In terms of the channel isospin cross section $\sigma_{if}$
with $i(f)=0$ or 1 being the initial and final isospins,
the $\Delta$ and $N^*(1440)$ production cross sections are given by
\begin{eqnarray}
&&\sigma(p+p \rightarrow n+\Delta^{++})
=\sigma_{10}+\frac{1}{2}\sigma_{11},\\
&&\sigma(p+p \rightarrow p+\Delta^{+})
=\frac{3}{2}\sigma_{11},\\
&&\sigma(n+p \rightarrow p+\Delta^{0})
=\frac{1}{2}\sigma_{11}+\frac{1}{4}\sigma_{10},\\
&&\sigma(n+p \rightarrow n+\Delta^{+})
=\frac{1}{2}\sigma_{11}+\frac{1}{4}\sigma_{10},\\
&&\sigma(p+p \rightarrow p+N^{\ast +})=0,\\
&&\sigma(n+p \rightarrow p+N^{\ast 0})
=\frac{3}{4}\sigma_{01},\\
&&\sigma(n+p \rightarrow n+N^{\ast +})
=\frac{3}{4}\sigma_{01}.
\end{eqnarray}
In the above, $\sigma_{if}(\sqrt{s})$ is defined as
\begin{equation}
\sigma_{if}(\sqrt{s})=\frac{\pi (\hbar c)^{2}}{2p^{2}}\alpha
(\frac{p_{r}}{p_{0}})^{\beta}\frac{m_{0}^{2}\Gamma^{2}(q/q_{0})^{3}}
{(s^{\ast}-m_{0}^{2})^{2}+m_{0}^{2}\Gamma^{2}},
\end{equation}
where the parameters $\alpha, \beta, m_{0}$ and $\Gamma$ are listed
in table \ref{isospin}.
Other quantities in the parameterization are defined as
in ref.\ \cite{verwest}.
The cross section for the production of a $N^{*}(1535)$
resonance is estimated from the empirical $\eta$ production cross section
according to\cite{wolf}
\begin{eqnarray}
&&\sigma(NN\rightarrow NN^{*}(1535))\approx
2\sigma(NN\rightarrow NN\eta),\\
&&\sigma(pn\rightarrow pn\eta)\approx
3\sigma(pp\rightarrow pp\eta)=3\frac{0.102(\sqrt{s}-2.424)}
{0.058+(\sqrt{s}-2.424)^2}~(mb).
\end{eqnarray}

Cross sections for double resonance production in processes shown in Eq. (2)
are estimated by subtracting from the measured inclusive $2\pi$ production
cross section in the nucleon-nucleon collision
\cite{data1} the contribution from $NN\rightarrow NN\rho$
and the $2\pi$ decay of the $N^{*}(1440)$ resonance in the
$NN\rightarrow NN^{*}(1440)$ reaction.
For example, we have for a $pp$ collision,
\begin{eqnarray}
&&\sigma(pp\rightarrow \sum_{I_{\Delta}I_{\Delta}}\Delta\Delta
+\sum_{I_{\Delta}I_{N^{*}}}
\Delta N^{*}(1440)) \nonumber \\
&&=4\sigma(pp\rightarrow pp\pi^{+}\pi^{-})
-2\sigma(pp\rightarrow pp\rho^{0})-0.4\sigma(pp\rightarrow NN^{*}(1440).
\end{eqnarray}
We show in Fig.\ 1 the experimental $pp$ inelastic
cross section (solid line) together with its decomposition
into contributions
from the production of various resonances.  It is seen that
for $\sqrt{s}\geq 3.0$ GeV the contribution from double resonance
production, shown by the dotted line, is about one third of the
total inelastic cross section.
We have assumed that the double resonance
production cross sections are the same for all channels
allowed by charge conservation. This approximation is
strongly supported by recent theoretical studies
based on the one-boson-exchange model\cite{aichelin}.

The cross sections for $\rho$ and $\omega$ production in reactions shown
in Eq. (3) are taken directly from the experimental data\cite{data1}.
As an example, we show in Fig.\ 2 the experimental cross sections for the
reactions $pp\rightarrow pp\omega$ and $pp\rightarrow pp\rho^0$.
The cross section for quasi-$3\pi$ production
$NN\rightarrow \Delta\Delta\pi$
is taken as the difference between the measured
inclusive $3\pi$ and $\omega$
production cross sections. For a $pp$ collision, we have
\begin{equation}
\sigma(pp\rightarrow \sum_{I_{\Delta}I_{\pi}}\Delta\Delta\pi)\equiv
3[\sigma(pp\rightarrow pp\pi^{+}\pi^{-}\pi^0)+
\sigma(pp\rightarrow pn\pi^{+}\pi^{+}\pi^-)]-
\sigma(pp\rightarrow pp\omega).
\end{equation}
It is seen from Fig.\ 1 that quasi-$3\pi$ production dominates the
inelastic cross section in the energy region
$\sqrt{s}\geq 3.5$ {\rm GeV} which corresponds
to the c.m. energies of colliding nucleon pairs in the
early stage of heavy-ion collisions at AGS energies.
We then attribute the difference between
the experimental total nucleon-nucleon inelastic cross
section and the sum of cross sections from reactions shown in
Eqs. (1) to (3) as well as the kaon production
cross sections from reactions shown later in Eq.\ (\ref{kaon})
to the quasi-$4\pi$ production process
$NN\rightarrow\Delta\Delta\rho$ (thin solid line in Fig.\ 1).
In this way, the total inelasticity of the nucleon-nucleon
collision is properly treated as only a limited, though large,
number of reactions have been incorporated.
The errors introduced by this approximation
for the quasi-$4\pi$ production cross section is small.
{}From Fig.\ 1, it is seen that the maximum cross section for the process
$NN\rightarrow \Delta\Delta\rho$ at $(\sqrt{s})_{NN}=4.9$ GeV,
corresponding to the c.m. energy of colliding nucleon pairs at
a beam momentum of 11.6 GeV/c, is about 3.0 mb, which,
however, has to be compared with the 13, 9 and 5 mb for the final
states $\Delta\Delta\pi, \Delta\Delta+\Delta N^{*}$ and $N\Delta+NN^{*}$,
respectively. Cross sections for reactions shown in  Eqs. (5) to (7) are taken
to be the same as in the nucleon-nucleon collision
having the same center-of-mass energy and total charge.

Using the rejection method,
the masses of $\Delta,~N^{*}(1440), ~N^{*}(1535)$ and $\rho$
are distributed according to the modified single or joint
Breit-Wigner distributions with momentum-dependent widths.
For single resonance production, the cross section
can be expressed in terms of the transition matrix
element ${\cal M}_{NN'\rightarrow N''r}$
\begin{equation}
\label{1dpro}
\sigma(NN'\rightarrow N''r)=\frac{m_{r0}^2 m_{N}^{3}}{\pi s p_{i}}\cdot
\int_{m_{N}+m_{\pi}}
^{\sqrt{s}-m_{N}}\frac{dm_{r}}{2\pi}P_1(m_{r})\cdot\int
\frac{d\Omega}{4\pi}\sum_{s_is_f}|{\cal M}_{NN'\rightarrow N''r}|^{2},
\end{equation}
where $P_1(m_{r})$ is the modified Breit-Wigner function
\begin{equation}
P_1(m_{r})=\frac{p_f\cdot m_{r}\cdot \Gamma(m_{r})}{(m_{r}^{2}
-m_{r0}^{2})^{2}+m_{r0}^{2}\Gamma_{r}^{2}(m_{r})}.
\end{equation}
In the above, $p_i$ and $p_f$ are the nucleon momenta in the
center-of-mass frame of $NN'$ and $N''r$, respectively.
The centroid and width of the resonance are denoted by $m_{r0}$ and
$\Gamma(m_{r})$, respectively.  As discussed in detail
in refs.\ \cite{dani91,li93}, to obtain an accurate resonance mass
distribution requires knowledge about the mass dependence of the matrix
element.  For simplicity, we assume that the matrix element is independent
of the mass, and the mass distribution of the resonance is then given
by $P_1(m_r)$. The momentum
factor $p_f$ in $P_1(m_r)$ plays an important role in collisions
at energies near the threshold for resonance production as it
suppresses the production of resonances with masses near
the maximum mass $m_{max}=\sqrt{s}-m_n$.

For double resonance production $NN'\rightarrow r_1r_2$, one has
\begin{equation}
\label{2dpro}
\sigma(NN'\rightarrow r_1r_2)=\frac{4m_n^2m_{r10}^2m_{r20}^2}
{\pi s p_{i}(1+\delta_{r_1r_2})}\cdot
\int\int\frac{dm_{r1}dm_{r2}}{4\pi^2}P_2(m_{r1},m_{r2})\cdot\int
\frac{d\Omega}{4\pi}\sum_{s_is_f}|{\cal M}_{NN'\rightarrow r_1r_2}|^{2},
\end{equation}
where $P_2(m_{r1},m_{r2})$ is the joint Breit-Wigner function
\begin{equation}\label{2mass}
P_2(m_{r1},m_{r2})=\frac{p_f\cdot m_{r1}\cdot
\Gamma_1}{[(m_{r1}^{2}-m_{r10}^2)^2
+m_{r10}^2\Gamma_1^2]}\cdot
\frac{m_{r2}\cdot \Gamma_2}{[(m_{r2}^{2}-m_{r20}^2)^2
+m_{r20}^2\Gamma_2^2]},
\end{equation}
with the conditions
\begin{eqnarray}
\label{limit}
m_n+m_{\pi}\leq &m_{r1}&
\leq \sqrt{s}-m_{r2},\nonumber \\
m_n+m_{\pi}\leq &m_{r2}&\leq\sqrt{s}-m_n-m_{\pi}.
\end{eqnarray}
In Eq.\ (\ref{2mass}), $p_f$ is the momentum of the resonances in their
center-of-mass frame.   Unlike other models for heavy-ion collisions
at AGS energies, resonances in our model can be produced
with a large range of masses.

For reactions producing a $\rho$ resonance, one has the similar
expression for the production cross section. The $\rho$ mass distribution
is
\begin{equation}
P(m_{\rho})=\frac{\Gamma^2(m_{\rho})}{(m_{\rho}-0.77)^{2}
+0.25\Gamma^2(m_{\rho})}.
\end{equation}

Moreover, in calculating the masses, decay probabilities and formation
cross sections of baryon resonances,
the following momentum-dependent widths are used,
\begin{eqnarray}
\Gamma(\Delta(1232)) &=& 0.47 q^{3}/
             \left(m_{\pi}^{2}+0.6q^{2}\right)~{\rm (GeV)},\\\
\Gamma(N^{*}(1440))&=&0.20(q/q_0)^{3}~{\rm (GeV)},\\\
\Gamma(N^{*}(1535))&=&0.15(q/q_0)~{\rm (GeV)},
\end{eqnarray}
where $q_{0}$ is the nucleon momentum in
the rest frame of the resonance with its mass at the centroid.
However, constant widths of 0.151 and 0.0084 {\rm GeV} are
used for $\rho$ and $\omega$, respectively.

The reverse reactions should be treated on the same footing.
However, in the present version of the model,
we have limited ourselves to only two-body collisions as indicated
by the arrows in Eqs.\ (1-7). In view of the high baryon density reached
in heavy-ion collisions at AGS energies, many-body interactions should
become increasingly important as the compression gradually
increases during the collision.
Both the physics and techniques of treating
many-body interactions are currently of great intreat but
unfortunately without clear solution. We therefore postpone the
inclusion of three-body and possibly many-body collisions to a later
version of the model. Cross sections for the two-body reverse reactions
are calculated by the detailed balance relation taking into account the
finite widths of the resonances.
For the absorption of a single resonance $N''r\rightarrow NN'$
we have the cross section
\begin{equation}
\label{1dabs}
\sigma(N''r\rightarrow NN')=\frac{m_{r} m_{N}^{3}p_i}
{2g_r\pi s p_{f}(1+\delta_{NN'})}\int
\frac{d\Omega}{4\pi}\sum_{s_is_f}|{\cal M}_{N''r\rightarrow NN'}|^{2},
\end{equation}
where $g_r=4(2)$ is the spin degeneracy of the resonance $r=\Delta(N^{*})$.
The factor $(1+\delta_{NN'})$ takes into account the case of having
two identical nucleons in the final state.
Using $|{\cal M}_{N''r\rightarrow NN'}|^{2}=
|{\cal M}_{NN'\rightarrow N''r}|^{2}$, Eqs.\ (\ref{1dpro})
and (\ref{1dabs}) lead to the following
detailed balance relation \cite{dani91}
\begin{eqnarray}
\label{det1}
\sigma(N''r\rightarrow NN')&=&\frac{m_r}{2g_{r}m_{r0}^2}
\cdot\frac{1}{(1+\delta_{NN'})}
\cdot\frac{p_{i}^{2}}{p_f}\nonumber \\
&\cdot&\sigma(NN'\rightarrow N''r)
\cdot\left(\int_{m_{\pi}+m_{N}}^{\sqrt{s}-m_{N}}\frac{dm'_{r}}{2\pi}
P_1(m'_{r})\right)^{-1}.
\end{eqnarray}

Similarly, for the absorption of two resonances $r_1r_2\rightarrow NN'$
we have
\begin{equation}
\label{2dabs}
\sigma(r_1r_2\rightarrow NN')=\frac{m_{r1}m_{r2} m_{N}^{2}p_i}
{g_{r1}g_{r2}\pi s (1+\delta_{NN'})p_f}\int
\frac{d\Omega}{4\pi}\sum_{s_is_f}|{\cal M}_{r_1r_2\rightarrow NN'}|^{2}.
\end{equation}
The detailed balance relation in this case then reads as
\begin{eqnarray}\label{det2}
\sigma(r_1r_2\rightarrow NN')&=&\frac{m_{r1}m_{r2}}
{4g_{r_1}g_{r_2}m_{r10}^2m_{r20}^2}\cdot
\frac{1+\delta_{r_1r_2}}{1+\delta_{NN'}}
\cdot\frac{p_{i}^{2}}{p_f}\nonumber \\
&\cdot&\sigma(NN'\rightarrow r_1r_2)
\cdot\left(\int\frac{dm'_{r1}}{2\pi}\frac{dm'_{r2}}{2\pi}
P_2(m'_{r1},m'_{r2})\right)^{-1}.
\end{eqnarray}

Limits of the mass integration are given by Eq.\ (\ref{limit}).
In the case, $P_1(m_{r})=\pi p_f
\delta(m_r-m_{r0})/(2m_{r0})$ and $\sqrt{s}\rightarrow \infty$,
the above relations reduce to the standard detailed
balance relations for the production and absorption of resonances with
fixed masses
\begin{eqnarray}
\label{d1}
\sigma(N^{''}r\rightarrow NN')=\frac{2}{g_{r}}\cdot\frac{1}
{1+\delta_{NN'}}
\cdot \frac{p_{i}^{2}}{p_{f}^{2}}\cdot \sigma(NN'\rightarrow N''r),\\
\sigma(r_1r_2\rightarrow NN')=\frac{4}{g_{r1}g_{r2}}\cdot
\frac{1+\delta_{r_1r_2}}
{1+\delta_{NN'}}
\cdot \frac{p_{i}^{2}}{p_{f}^{2}}\cdot \sigma(NN'\rightarrow r_1r_2).
\end{eqnarray}
At AGS energies, resonances
with large widths can be produced, especially in the early
stage of the collisions, it is then necessary to use the
detailed balance relations of Eqs.\ (\ref{det1}) and (\ref{det2})
to treat their annihilation.

\subsection{Baryon-meson interactions}
One can also separate meson-baryon interactions into the
elastic and inelastic parts. For the elastic interaction, we
treat them via the formation of baryon resonances,
\begin{eqnarray}\label{fres}
\pi N&&\leftrightarrow \Delta, N^{*}(1440), N^{*}(1535),\\
\eta &&N\leftrightarrow N^{*}(1535),
\end{eqnarray}
as well as direct reactions,
\begin{eqnarray}
&&\pi+N(\Delta, N^{*})\rightarrow \pi+N(\Delta,N^{*}),\\
&&\rho+N(\Delta, N^{*})\rightarrow \rho+N(\Delta, N^*),\\
&&K^++N(\Delta, N^*)\rightarrow K^++N(\Delta, N^*).
\end{eqnarray}

The standard Breit-Wigner form\cite{pilkuhn} of resonance
formation in meson-nucleon interactions can be rewritten as
\begin{equation}
\sigma(M+N)=\sigma_{\rm max}\cdot (\frac{q_{0}}{q})^{2}\cdot
\frac{\frac{1}{4}\Gamma^2(m_r)}{(m_r-m_{r0})^2+\frac{1}{4}\Gamma^2(m_r)},
\end{equation}
where $q_{0}$ is the meson momentum at the centroid $m_{r0}$
of the resonance mass distribution. The mass $m_r$ of the
produced baryon resonance
is uniquely determined by reaction kinematics.
The maximum cross sections are given by
\begin{eqnarray}
\sigma_{\rm max}(\pi^{+}+p\rightarrow\Delta^{++})=&\sigma_{\rm max}(\pi^{-}+n
      \rightarrow\Delta^{-})&=190\ {\rm mb},\\
\sigma_{\rm max}(\pi^{0}+p\rightarrow\Delta^{+})=&\sigma_{\rm max}(\pi^{0}+n
      \rightarrow\Delta^{0})&=50\ {\rm mb},\\
\sigma_{\rm max}(\pi^{-}+p\rightarrow\Delta^{0})=&\sigma_{\rm max}(\pi^{+}+n
      \rightarrow\Delta^{+})&=\ 30\ {\rm mb},\\
\sigma_{\rm max}(\pi^{-}+p\rightarrow {\rm N}^{*0}(1440))
=&\sigma_{\rm max}(\pi^{0}+n\rightarrow {\rm N}^{*0}(1440))&
=\ 6 \ {\rm mb},\\
\sigma_{\rm max}(\pi^{+}+n\rightarrow {\rm N}^{*+}(1440))
=&\sigma_{\rm max}(\pi^{0}+p
\rightarrow {\rm N}^{*+}(1440))&=\ 12\ {\rm mb},\\
\sigma_{\rm max}(\pi^{-}+p\rightarrow {\rm N}^{*0}(1535))
=&\sigma_{\rm max}(\pi^{0}+n\rightarrow {\rm N}^{*0}(1535))&
=\ 8\ {\rm mb},\\
\sigma_{\rm max}(\pi^{+}+n\rightarrow {\rm N}^{*+}(1535))
=&\sigma_{\rm max}(\pi^{0}+p
\rightarrow {\rm N}^{*+}(1535))&=\ 16\ {\rm mb},\\
\sigma_{\rm max}(\eta+p\rightarrow {\rm N}^{*+}(1535))
=&\sigma_{\rm max}(\eta+n\rightarrow {\rm N}^{*0}(1535))&
=\ 74\ {\rm mb}.
\end{eqnarray}
In evaluating $\sigma_{\rm max}$, we have taken into account
properly the decay branching ratios of the $N^*(1440)$ and $N^*(1535)$
resonances.

The formation of the three baryon resonances in the reactions shown in
Eq.\ (\ref{fres}) accounts almost entirely the $\pi+N$ elastic cross
sections at low energies.
This is demonstrated in Fig.\ 3 and Fig.\ 4 where the
experimental cross sections for the elastic scattering of $\pi^-+p$
and $\pi^++p$ are compared with the sum of the three baryon resonance
formation cross sections. It is also seen from the figures that the
formation of the $\Delta$ resonance alone is not enough to describe
even the low energy part of $\pi+N$ scattering. At higher
energies, the elastic cross section is mainly due to the formation
of higher resonances which are not included in the present version of the
model. We therefore attribute the difference between the experimental elastic
cross section and the contribution from the three lowest baryon resonances
to the direct process $\pi+N\rightarrow \pi+N$. The contribution
from direct $\pi+N$ scattering is shown by the dotted lines
in Fig.\ 3 and Fig.\ 4.

The experimentally unknown cross sections, such as
those for $\pi^{0}+N$, $\pi+\Delta (N^{*})$ and
$\rho+N(\Delta, N^*)$ reactions, are calculated by
using a resonance model that includes
heavier baryon resonances with masses up to about 2.0 GeV.
Neglecting interferences between resonances, one has
\begin{equation}\label{highr}
\sigma(M+B)=1.3\frac{\pi}{k^{2}}\sum_{R}\frac{(2J_R+1)}{(2S_M+1)(2S_B+1)}
\frac{\Gamma^2_R(M+B)}{(\sqrt{s}-m_R)^2
+0.25\Gamma_{R}^2(total)}.
\end{equation}
The pre-factor 1.3 is introduced so that the
high energy part of $\pi^{+}+p$ data can be fitted
and is mainly due to the neglect of interferences.
The summation is over all baryon resonances with masses up to 2 GeV.
The total width $\Gamma_R(total)$ and partial widths $\Gamma_R(M+B)$
of heavier $N^{*}$ and $\Delta$ resonances used in the summation are listed
in Table \ref{nres} and Table \ref{dres}, respectively.
For illustrations, we show in Fig.\ 5 and Fig.\ 6 the calculated
$\pi^{0}+p(n)$ and $\pi^{+}+\Delta^-(\Delta^0)$ elastic cross sections.
For the $K^++N (\Delta, N^*)$ scattering a constant cross section of 10 mb is
used according to the available data.

The decay of resonances, such as
$\Delta (N^{*}(1440),~N^{*}(1535))\rightarrow \pi+N$,
$N^{*}(1440)\rightarrow 2\pi+N$, $N^{*}(1535)\rightarrow \eta+N$,
$\rho\rightarrow 2\pi,$ and $\omega\rightarrow 3\pi$,
during each time step $dt$ are treated by
the Monte Carlo method. The decay probability of a resonance is
calculated via
\begin{equation}
P_{decay}=1-exp[-dt\cdot\Gamma(m_r)/(\gamma\cdot\hbar)],
\end{equation}
where $\gamma=E/m_r$ is the Lorentz factor associated with the moving
resonance.
For the decay of $N^{*}(1440)$ and $N^{*}(1535)$
resonances the actual final state is chosen according to the
corresponding branching ratios. Namely,
$B(N^{*}(1440)\rightarrow \pi+N)=0.6,~
B(N^{*}(1440)\rightarrow 2\pi+N)=0.4
$ for $m(N^*(1440))\geq 2m_n+m_{\pi}$,
otherwise $B(N^*(1440)\rightarrow \pi+N)=1$.
For the $N^*(1535)$ resonance we use
$B(N^{*}(1535)\rightarrow \pi+N)\approx B(N^{*}(1535)\rightarrow
\eta+N)=0.5$. Finally, $\Lambda$ and $\Sigma$ are allowed to decay
into a nucleon and a pion only at the very end of the reaction.
It is worthy to mention that the Bose-Einstein
enhancement factor $(1+f_{\pi})$ for a pion in the final state
has not been included in meson+nucleon
collisions and decays of resonances. This is certainly another
aspect to be improved in a later version of the model in view of the
high pion densities reached in heavy-ion collisions at AGS energies.

The $\pi+N$ inelastic collision starts at
$(\sqrt{s})_{\pi N}\approx 1.2$ {\rm GeV} due to the $\pi\pi N$ final state.
As the energy increases the number of final state quickly increases to
the order of 200. Most inelastic reactions lead to pion and kaon production.
The final states can also consist of
baryon and meson resonances. To treat the inelastic reactions
in a numerically tractable way, we
model them through the production of $\Delta,~\rho $ and $\omega$
resonances, i.e.,
\begin{eqnarray}
&&\pi+N\leftrightarrow \Delta+\pi,\\
&&\pi+N\leftrightarrow \Delta+\rho,\\
&&\pi+N\leftrightarrow \Delta+\omega,
\end{eqnarray}
and those producing kaons shown later in Eq.\ (\ref{kaon1}).
We have thus included only final
states with at most 4 quasipions. Again, the above decomposition
of the cross sections involves
some uncertainties and approximations. However, as we will discuss
later our final results are not sensitive to them
due to the large number of final-state interactions.

For two and three pion production in
$\pi+N$ collisions, we assume that they are mainly through
the production of $\Delta\pi$ and $\Delta\rho$, respectively.
For example, for the $\Delta\pi$ final state in
the $\pi^++p$ interaction, we use
\begin{eqnarray}
\sigma(\pi^+p\rightarrow \sum_{I_{\Delta}I_{\pi}}\Delta\pi)&\equiv&
\sigma(\pi^+p\rightarrow p\pi^+\pi^0)\nonumber \\
&+&\sigma(\pi^+p\rightarrow n\pi^+\pi^+)\nonumber \\
&+&\sigma(\pi^+p\rightarrow p\rho^+)\nonumber \\
&+&2\sigma(\pi^+p\rightarrow \Delta^{++}\pi^0),
\end{eqnarray}
and for the $\Delta\rho$ final state, we use
\begin{eqnarray}
\sigma(\pi^+p\rightarrow \sum_{I_{\Delta}I_{\rho}}\Delta\rho)&\equiv&
3\sigma(\pi^+p\rightarrow p\pi^+\pi^+\pi^-)\nonumber \\
&+&3\sigma(\pi^+p\rightarrow p\pi^+\rho^0)\nonumber \\
&+&2\sigma(\pi^+p\rightarrow \Delta^{++}\rho^0)\nonumber \\
&+&\sigma(\pi^+p\rightarrow p\omega).
\end{eqnarray}
The difference between the
experimental total $\pi+N$ inelastic cross section and the
cross sections for producing $\Delta\pi$, $\Delta\rho$ and
kaons are then attributed to the production of $\Delta\omega$, i.e.,
\begin{eqnarray}
\sigma(\pi^{+}p\rightarrow \Delta^+\omega)&\equiv&
\sigma^{inel.}_{exp.}(\pi^{+}p)-\sigma(\pi^{+}p\rightarrow
\sum_{I_{\Delta}I_{\pi}}\Delta\pi)\nonumber \\
&-&\sigma(\pi^+p\rightarrow \sum_{I_{\Delta}I_{\rho}}\Delta\rho)
-\sigma(\pi^+p\rightarrow K^+X).
\end{eqnarray}
As an example, the decomposition of the inelastic
$\pi^++p$ cross section is shown in Fig.\ 7.
It is seen that the cross sections for having 2, 3 and 4
quasipions are comparable at $\sqrt{s}\approx 2.5 $ {\rm GeV}, but at higher
energies the $4\pi$ channel dominates. As we shall see later,
the distribution of meson+baryon
center of mass energies peaks at about $\sqrt{s}\approx 1.5 $
{\rm GeV} in the reaction of Au+Au at $P_{beam}/A$= 11.6 {\rm GeV/c} and
falls off very quickly towards $\sqrt{s}\approx 2.5 $ {\rm GeV}.
It thus makes us feel rather confident that the errors introduced
in the quasi-$4\pi$ channel in our previous decomposition is very small.
The two-body reverse reactions of the $\pi+N$ inelastic collisions are
similarly treated as for the two-body reverse reactions of the inelastic
baryon-baryon collisions.

\subsection{Meson-meson interactions}
We model pion-pion elastic collisions through the formation of
a $\rho$ meson, i.e., $\pi+\pi\leftrightarrow \rho$, and the direct
process $\pi+\pi\rightarrow \pi+\pi$.
The latter takes into account the case when the quantum numbers
of colliding pions forbid the formation of a $P_{11}$ meson $\rho$.

In terms of the channel isospin cross sections $\sigma^{I}$
with $I$=0, 1, and 2,
one finds using the Glebsch-Gordan coefficients the following isospin
decomposition of $\pi\pi$ scattering cross sections
\begin{eqnarray}
&&\sigma(\pi^+\pi^+)=\sigma(\pi^-\pi^-)=\sigma^2,\\
&&\sigma(\pi^+\pi^0)=\sigma(\pi^-\pi^0)=\frac{1}{2}\sigma^2
+\frac{1}{2}\sigma^1,\\
&&\sigma(\pi^+\pi^-)=\frac{1}{6}\sigma^2
+\frac{1}{2}\sigma^1+\frac{1}{3}\sigma^0,\\
&&\sigma(\pi^0\pi^0)=\frac{1}{3}\sigma^0+\frac{2}{3}\sigma^2,\\
&&\sigma(\pi^+\pi^-\leftrightarrow \pi^0\pi^0)=-\frac{1}{3}\sigma^0
+\frac{1}{3}\sigma^2.
\end{eqnarray}
Because of the symmetrization of two-pion states, only even
partial waves are allowed for $\sigma^0$ and $\sigma^2$ and
odd partial waves for
$\sigma^1$. Low energy $\pi\pi$ scattering are therefore described
by the three partial-wave cross sections $\sigma^0_0, \sigma^{2}_0$
and $\sigma^1_1$. The formation of a $\rho$ resonance
has thus the probability
\begin{equation}
P_{\rho}(\pi\pi)=\frac{1}{2}\frac{\sigma^{1}_1}{\sigma(\pi\pi)},
\end{equation}
while the direct $\pi\pi\to\pi\pi$ reaction
has the probability of $1-P_{\rho}(\pi\pi)$.

The partial wave cross sections $\sigma^I_L$
are related to the phase shift $\delta^I_L$ via
\begin{equation}
\sigma^I_L=\frac{8\pi}{q^2}(2L+1)sin^2\delta_L^I,
\end{equation}
where $q$ is the momentum of the pion in the center-of-mass of
the two pions.  For $\delta_L^I$ we use the parameterization of
ref.\ \cite{bert}
\begin{eqnarray}
\delta^0_0&&=tan^-1\left(\frac{1.03q}{5.8m_{\pi}-\sqrt{s}}\right),\\
\delta^1_1&&=tan^{-1}\left(\frac{\Gamma_{\rho}/2}{0.77-\sqrt{s}}\right),\\
\delta^{2}_0&&=-0.12q/m_{\pi},
\end{eqnarray}
where $\Gamma_{\rho}=0.095q[q/m_{\pi}/(1+(q/0.77)^2)]^2$.
The isospin-averaged cross section for $\pi\pi$ scattering
\begin{equation}
\bar{\sigma}(\pi\pi)=\frac{1}{9}\sigma^0_0+\frac{1}{3}\sigma^1_1+
\frac{5}{9}\sigma^2_0
\end{equation}
is shown in the upper window of Fig. 8
as a function of the center-of-mass energy $\sqrt{s}$.
Since the densities of $\rho$ and $\omega$ are rather small at AGS energies,
we have neglected in the present version of the model the
elastic scattering of $\pi+\rho, \pi+\omega$
and $\rho+\omega$.

The inelastic collisions among mesons are modeled through
the reaction, $MM\rightarrow K\bar{K}$. The
cross section for this reaction is
highly uncertain, here we use the cross section calculated from the
$K^{*}$-exchange model of ref.\ \cite{ko}. The lower window of
Fig.\ 8 shows the isospin-averaged cross section for
the reaction $\pi\pi\rightarrow K\bar{K}$.
Calculated cross sections in ref. \cite{ko} for the reactions
$\rho\rho\rightarrow K\bar{K}$ and
$\pi\rho\rightarrow K\bar{K}$ have values about 0.3 mb, except
at energies very close to the threshold, which we shall
use in our model.  Furthermore, we assume that
for the reactions $\pi\omega\rightarrow K\bar{K}$ and
$\rho\omega\rightarrow K\bar{K}$, the cross sections have similar values.

\subsection{$K^+$ production in baryon-baryon and baryon-meson interactions}
We are at present mainly interested in
$K^+$ production. The following kaon production channels in
baryon-baryon collisions are included
\begin{eqnarray}\label{kaon}
&&NN\rightarrow N\Lambda(\Sigma)K,~\Delta\Lambda(\Sigma)K,\\
&&NR\rightarrow N\Lambda(\Sigma)K,~\Delta\Lambda(\Sigma)K,\\
&&RR\rightarrow N\Lambda(\Sigma)K,~\Delta\Lambda(\Sigma)K,\label{kaon2}
\end{eqnarray}
where R denotes a $\Delta,~N^{*}(1440)$ or $N^{*}(1535)$.
We use the approximation that the kaon production cross sections
in reactions induced by resonances are the same as in nucleon-nucleon
collisions at the same center-of-mass energy which are mainly taken
from the data compilation of refs.\ \cite{data1,lbl}.
Since the experimental data on kaon production
are very limited, we use only the isospin-averaged cross sections.
Based on the one-pion exchange model these cross sections can
be expressed in terms of the experimentally known ones \cite{rand}
\begin{eqnarray}
&&\bar{\sigma}(NN\rightarrow N\Lambda K^+)\approx\bar{\sigma}
(NN\rightarrow \Delta\Lambda K^+)
\approx\frac{3}{2}\sigma(pp\rightarrow p\Lambda K^+),\\
&&\bar{\sigma}(NN\rightarrow N\Sigma K^+)\approx\bar{\sigma}
(NN\rightarrow \Delta\Sigma K^+)
\approx\frac{3}{2}[\sigma(pp\rightarrow p\Sigma^0 K^+)
+\sigma(pp\rightarrow p\Sigma^+ K^0)].
\end{eqnarray}
The threshold energies are 2.56, 2.74, 2.63 and 2.77 {\rm GeV}
for the four final states $N\Lambda K, \Delta \Lambda K, N\Sigma K$
and $\Delta \Sigma K$, respectively. The actual final state of a given
collision is determined using the Monte Carlo method
according to the relative ratios among these cross sections.

The kaon momentum distribution from a baryon-baryon interaction was
parameterized by Randrup and Ko according to a modified
phase space\cite{rand}, i.e.,
\begin{equation}\label{nn}
{E\over p^2}{d^3\sigma (\sqrt s)\over dpd\Omega}=
\sigma_{K^+}(\sqrt{s}){E\over 4\pi p^2}{12\over p_{\rm max}}
\left (1-{p\over p_{\rm max}}\right )\left ({p\over p_{\rm max}}\right )^2,
\end{equation}
where $p_{\rm max}$ is the maximum kaon momentum given by
\begin{equation}
p_{\rm max}={1\over 2}\sqrt {\big[ s-(m_B+m_Y+m_K)^2\big]
\big[ s-(m_B+m_Y-m_K)^2\big] /s}.
\end{equation}
The angular distribution is taken as isotropic in the center-of-mass frame
of colliding baryons.

In meson-baryon interactions, kaons are produced through
\begin{eqnarray}\label{kaon1}
&&\pi+N(\Delta, N^{*})\rightarrow \Lambda (\Sigma)+K,\\
&&\rho+N(\Delta, N^{*})\rightarrow \Lambda (\Sigma)+K,\\
&&\omega+N(\Delta, N^{*})\rightarrow \Lambda (\Sigma)+K.
\end{eqnarray}

For pion-nucleon collisions,
the isospin-averaged cross sections for kaon production
can be expressed in terms of the available data
\cite{cugnon}
\begin{eqnarray}
&&\bar{\sigma}(\pi N\rightarrow \Lambda K^+)
\approx\frac{1}{4}\sigma(\pi^+ n\rightarrow \Lambda K^+),\\
&&\bar{\sigma}(\pi N\rightarrow \Sigma K^+)
\approx\frac{1}{4}[\sigma(\pi^- p\rightarrow \Sigma^- K^+)
+\sigma(\pi^- p\rightarrow \Sigma^0 K^0)
+\sigma(\pi^+ p\rightarrow \Sigma^+ K^+)].
\end{eqnarray}
Since there are no data available for kaon
production in resonance induced reactions
(e.g., $\rho(\omega)+N(\Delta, N^{*})$ and $\pi+\Delta (N^*)$),
they are taken for simplicity to be the same as in the $\pi+N$ collision
at the same center-of-mass energy.
This is probably a resonable, minimal assumption one can make.
However, it is interesting to
note that theoretical efforts have recently been made to calculate
these unknown cross sections (e.g., \cite{fasselor}), and we plan to
incorporate these results in future versions of the ART model.

The experimental information about the angular distribution of the outgoing
kaon in a meson-baryon collision is rather sparse but
shows a complicated structure\cite{cugnon}. Here,
we use an isotropic distribution in the meson-baryon center of mass
frame. A test using a forward-backward peaked angular distribution in
the baryon-meson c.m. frame shows little
change in the final kaon distribution in heavy-ion collisions
as a result of kaon final-state interactions.

Finally, we would like to point out that in reactions $RN\to NYK$ and
$RR\to NYK$, which are dominated by one-pion exchange, the pion can be
on-shell. The latter contribution is equivalent to
a two-step process, i.e., a resonance decaying into a physical
pion and a nucleon, and the subsequent production of a kaon from the
pion-baryon interaction.  To avoid double counting,
one can follow the approach of Ref. \cite{liko} by including only the
reactions $RN\to NYK$ and $RR\to NYK$ but not the reaction $\pi B\to NYK$.
This requires, however, a model for the reactions
$RN\to NYK$ and $RR\to NYK$. In the present paper, we use instead
the assumption that the cross sections for the reactions $RN\to NYK$
and $RR\to NYK$ due to off-shell pions are the same as
in $NN\to NYK$ reaction at the same center-of-mass energy,
which invloves only an off-shell pion.  Then, we should include both the
reactions $RN\to NYK$ and $RR\to NYK$ and the reactions $\pi B\to NYK$
in the transport model.

\subsection{Hadron momentum distributions}
In this section, we discuss the method we use
to determine the final particle momenta
in a hadron-hadron collision, which does not involve kaons,
In principle, to determine the momentum of a particle in
a multiparticle production process, one needs to know
the interaction matrix element which is model-dependent.
For simplicity, we determine the momentum distribution of the produced
particles according to the phase space.
Besides imposing the energy-momentum conservation, we also make use of the
systematics observed in multiparticle production processes.
These systematics are often given by empirical formula
that are fitted to the inclusive data.

An important feature of energetic hadron-hadron collisions is the
``leading'' particle behaviour. In our model this is ensured by
requiring the outgoing baryons to have the same or similar
identities as the incident ones so that their longitudinal
directions are retained when performing the momentum
transformation from the baryon-baryon
c.m. frame to the nucleus-nucleus c.m. frame.
Moreover, baryons in the final state
of an energetic hadron-hadron collision have a typical
forward-backward peaked
angular distribution and a soft transverse momentum distribution.
For baryon-baryon collisions with $\sqrt{s}\leq 3.0 $ {\rm GeV}
where single-resonance production dominates, we use the angular
distribution obtained from
fitting the $pp\rightarrow N\Delta$ data\cite{wolf90}, i.e.,
\begin{equation}
\frac{1}{\sigma}\frac{d\sigma}{d\Omega}=b_1(s)+3b_3(s)cos^2\theta,
\end{equation}
where $b_1(s)=0.5$ and $b_3(s)=0$ for $\sqrt{s}\leq 2.14$ {\rm GeV}, and
\begin{eqnarray}
&&b_1(s)=29.03-23.75\sqrt{s}+4.87s,\\
&&b_3(s)=-30.33+25.53\sqrt{s}-5.30s,
\end{eqnarray}
for $2.14\leq \sqrt{s}\leq 2.4 $ {\rm GeV}, while for
$2.4 \leq \sqrt{s}\leq 3.0$
{\rm GeV}, $b_1(s)=0.06, b_3(s)=0.4$.
This distribution is shown in Fig.\ 9, and
it is seen that the forward-backward
peaked angular distribution becomes almost energy independent for
$\sqrt{s}\geq 2.3$ {\rm GeV}. This tendency is consistent with the systematics
found in $pp$ collisions at high energies where the inclusive proton
spectra can be well described by \cite{data2}
\begin{equation}\label{dis1}
\frac{d^{2}N}{d^3\vec{p^*}}=\frac{1.39}{\sqrt{s}}(1.+0.43x^{*}-0.84x^{*^2})
(e^{-3.78p_{t}^{2}}+0.47e^{-3.6p_{t}}),
\end{equation}
with $x^{*}$ being the scaled longitudinal momentum in the c.m. frame, i.e.,
$x^{*}=2p^*_z/\sqrt{s}$. The distribution has the properties of the naive
scaling, $p_t$ and $p^{*}_{z}$ factorization and the soft,
energy-independent transverse momentum distribution.

Here we adopt the same functional from of momentum distribution as given
by Eq.\ (\ref{dis1}) for all
baryons in collisions with $\sqrt{s}\geq 3.0$ {\rm GeV},
but slightly adjust the parameters in the transverse momentum
distribution so that we can reproduce the inclusive proton
momentum distribution from $pp$ collisions at
$P_{beam}=15$ {\rm GeV/c}.
Although both longitudinal and transverse momentum
distributions are needed for three-body final states, only
the transverse momentum distribution is required for two-body final states
as the longitudinal momentum can be determined
from the momentum conservation.
The momentum distribution determined by the
present model turns out to be very similar to Eq.\ (\ref{dis1}), i.e.,
\begin{equation}\label{md}
\frac{d^{2}N}{d^3\vec{p^*}}\propto (1.+0.5x^{*}-0.9x^{*^2})
(e^{-4p_{t}^{2}}+0.5e^{-10p_{t}}).
\end{equation}
In the upper window of Fig.\ 10, the proton transverse
momentum distribution from $pp$ collisions at
$p_{beam}=15$ Gev/c, which is used to determine the input momentum
distribution in Eq.\ (\ref{md}), is shown by the solid squares.
The solid line is a plot of the proton transverse momentum distribution
from Eq.\ (\ref{dis1}). The quality of the fit
is reasonably good although there is still some room
at low transverse momenta for further improvement.
The longitudinal momentum distribution
has a very similar behaviour. With the above transverse momentum distribution
for baryons and employing the assumption that
all resonances decay isotropically in their rest frames
we obtain the transverse momentum distribution of $\pi^-$ as shown
in the lower window of Fig.\ 10. The $\pi^{-}$ transverse momentum
distribution is then seen to fit reasonably well the empirical formula,
\begin{equation}
\frac{dN}{dp_t^2}(\pi^-)\propto (e^{-5.2p_{t}^{2}}+0.81e^{-4.3p_{t}}),
\end{equation}
of Ranft el al\cite{data2} for high energy $pp$ collisions.
For comparisons, we have also shown in Fig.\ 10 the generated
proton and $\pi^-$ transverse momentum distributions
from $pp$ collisions at $p_{beam}=3.0$ GeV/c. They, of course, do not
agree with the scaling formula valid at high energies.

The above procedure makes us feel confident about
our model for high energy heavy-ion collisions. We remark here, however,
that there is no direct, transparent extrapolation from $p+p$, $p+A$ to $AA$
collisions. The dynamics and the final momentum distributions of
hadrons in heavy-ion collisions are rather sensitive to the
modeling or prescription of individual
hadron-hadron scattering, besides the elementary momentum distributions
discussed above.

\subsection{Nuclear equation of state and causality}
The nuclear equation of state describes the response of the nuclear matter
to changes in excitation energies and densities. Current knowledge
on the nuclear equation of state is restricted to
a narrow region around the ground state. Extending
our knowledge about the nuclear equation of state to
different densities and/or excitation
energies has been a major goal of nuclear physics.

Mean-field effects have already been found to be important in
heavy-ion collisions at medium energies. At higher energies such as
at AGS, they have been ignored in ARC, for example.
We believe that the mean-field potential is not
negligible in heavy-ion collisions at AGS energies.
Although the forward scattering amplitudes
of hadron-hadron collisions in the high energy limit have been
found approximately purely imaginary\cite{wong}, the AGS energies
may not be high enough for the real part of the scattering
amplitude to completely vanish. Of course, the form and strength of
the mean field in hot dense hadronic matter is
uncertain and has been a subject of many discussions.
Secondly, although in the early stage of the collision
the kinetic energy is much higher than the potential
energy,  particles are gradually
slowed down and the mean field is expected to play
an increasingly important role
as the collision proceeds. In particular, the repulsive mean field in
the high density region tends to keep particles from coming close
and therefore reduces the maximum energy and baryon
densities reached in the collision should there be no mean field.
Moreover, in the expansion phase of the collision mean-field effects
become even stronger.
It is therefore essential to study how the collision dynamics is
affected by the nuclear mean field. Of course, the most
challenging task is to identify the experimentally observable
consequences of the nuclear mean field, so the information about
the nuclear equation of state can be obtained.

Without much reliable knowledge
about the equation of state in hot dense medium, we
use here the simple, Skyrme-type parameterization for the mean field
which has been widely used in heavy-ion collisions at and below
Bevalac energies,
\begin{equation}\label{field}
U(\rho)=a \frac{\rho}{\rho_{0}}+b(\frac{\rho}
{\rho_{0}})^{\sigma}+V_{Coulomb}.
\end{equation}
The corresponding energy per nucleon in nuclear matter at zero temperature
is given by
\begin{equation}
\frac{E}{A}=\frac{a}{2}(\frac{\rho}{\rho_0})
+\frac{b}{1+\sigma}(\frac{\rho}{\rho_0})^{\sigma}
+\frac{3}{5}E_{f}(\frac{\rho}
{\rho_{0}})^{\frac{2}{3}},
\end{equation}
where $E_{f}$=37.26 MeV is the Fermi energy.
The parameter $a$ is negative, $b$ is positive and $\sigma$ is larger than
one reflecting the fact that the nucleon-nucleon interaction has a
short-range repulsive part and a long-range attractive part. By
fitting to the ground state properties
of nuclear matter the three parameters are
determined in terms of the compressibility coefficient $K$ as
\begin{eqnarray}
a&=&-29.81-46.90\frac{K+44.73}{K-166.32} (MeV),\\
b&=&23.45\frac{K+255.78}{K-166.32} (MeV),\\
\sigma&=&\frac{K+44.73}{211.05}.
\end{eqnarray}
The compressibility coefficient $K$ can also be expressed in terms of
$a$, $b$, and $\sigma$, i.e.,
\begin{equation}
K = 9\rho (\partial P/\partial \rho)_{s}=9(\frac{p_{f}^{2}}{3m}+a+b\sigma).
\end{equation}

An important constraint on the nuclear equation of state in hot
dense medium is imposed by
the causality\cite{kapusta1,strotman}.
For nuclear matter at zero temperature the
adiabatic sound velocity is
\begin{equation}
V_s^2=\frac{1}{m}(\frac{\partial P}{\partial \rho})_s
=\frac{1}{m}[\frac{2}{3}E_f(\frac{\rho}{\rho_0})^{2/3}
+a\frac{\rho}{\rho_0}+b\sigma(\frac{\rho}{\rho_0})^{\sigma}].
\end{equation}
The causal requirement $V_s^2\leq c^2$ limits the density range applicable
for a given parameter set.
In Fig.\ 11, we show the adiabatic sound velocity and the energy per nucleon
as functions of density for the two equations of state corresponding to
$K$=377 MeV ($\sigma=$2, stiff) and
201 MeV ($\sigma=7/6$, soft). It is seen that the stiff and soft
equations of state violate the causality beyond
densities $\rho\geq 3\rho_0$ and $\rho\geq 7\rho_0$, respectively.
At finite temperatures the critical densities are reduced by less
than one density unit\cite{strotman}.
In the following, we shall use the soft equation
of state, and the stiff equation of station will only be mentioned
for the purpose of discussions.

The mean-field potentials for
baryon resonances are likely to be different from that of nucleons.
Resonance potentials have been studied previously \cite{ginocchio} and
are still less uncertain\cite{baldo} than the nucleon potential.
For example, significant efforts have been devoted to
understanding the effects of the
$\Delta$ potential on the nuclear matter saturation
properties\cite{baldo,betz,lee,haar}.
The $\Delta$ optical potential has also been studied in
charge-exchange reactions \cite{esbensen}, electron and $\gamma$
induced reactions\cite{connel} as well as $\pi$-nucleus
scattering \cite{horikawa,johnson}.
All studies indicate that the shape of the
$\Delta$ potential is very similar to the nucleon potential, probably a
little deeper than the nucleon potential. The exact form of the $\Delta$
potential somehow depends on the momentum and density.
Current theories give a large range of model parameters
for the $\Delta$ potential, and they often have conflicting features.
In view of the larger uncertainties
associated with resonance potentials, we assume here that
they are the same as the nucleon potential.

Since a large number of baryon resonances are produced in
heavy-ion collisions at AGS energies,
the reaction dynamics might be significantly
affected by the resonance potentials.
Relativistic heavy-ion collisions may
therefore provide valuable information about the
resonance potentials. However, this can only be possible if
other details of the collision dynamics are well understood.
The present work is thus also useful in this respect.

\section{Application of ART 1.0 to heavy-ion collisions at AGS energies}
We first discuss the range of beam energies in which
the model outlined above is applicable. In the energy range of
$E_{beam}/A\leq 3.0 $ {\rm GeV}/A, our model, as many other
nuclear transport models, is able to describe many aspects of
heavy-ion collisions. These include
the emission of light particles, the creation of pions, kaons and etas
as well as the transverse collective flow of various hadrons\cite{li93a,li93b}.
Since the particle production mechanism used in the model
is exclusively based on the excitation and decay of
several low-mass baryon and meson resonances,
the model is expected to be inadequate once the energy is above
about $E_{beam}/A=15$ GeV. At higher energies,
such as the SPS/CERN energies, heavier resonances and more importantly
the formation and fragmentation of strings and ropes are expected to
be the dominant mechanisms for particle production \cite{rqmd}.

The model ART 1.0 is designed intentionally to be most applicable
in heavy-ion collisions below the AGS energies.
As in other hadronic models, our model is
naturally limited only to interactions among hadrons. If
there were a phase transition to the QGP at AGS,
hadronic models, such as the present one, are still useful as their
predictions provide the background against which
new physical phenomena can be identified.
In fact, one of the motivations for the present work is to study whether
and how conditions for forming the
quark-gluon plasma can be achieved in heavy-ion collisions at beam energies
from 2 GeV/nucleon up to AGS energies.
Moreover, hadronic models are useful
for studying in-medium properties of hadrons. Theoretical studies
(e.\ g.\ \cite{brown,kaplan,hatsuda,asakawa}) have shown that the
properties of hadrons, such as the mass and
decay width, may be modified in hot dense hadronic matter
as a result of the partial restoration of chiral symmetry.
In the following, we shall apply the present model to study
several general, but important aspects
and issues of heavy-ion collisions at AGS energies.
In the subsequent work we plan to study among many other subjects: (i)
the effects of changing hadron properties in hot dense medium
on heavy-ion collision dynamics and, in particular, on the production
of exotic particles, (ii) the beam energy dependence of
the collision dynamics from 2 GeV/nucleon up to
AGS energies which will be carried out experimentally
by the EOS collaboration, and (iii) the role of quark and gluon
degrees of freedom at high densities and the study of physics
related to the phase transition to the QGP.

\subsection{Reaction dynamics at AGS energies}
To study the reaction dynamics of central heavy-ion collisions at AGS
energies, we shall first study in this section
the reaction rates of various hadron-hadron
collisions. To avoid the complexity introduced by the mean field,
we shall use the cascade mode of ART. In particular, we shall consider
head-on collisions of Au+Au
at $p_{beam}/A$=11.6 GeV/c as an example for discussions.

Fig.\ 12 shows the reaction rates of nucleon-nucleon scattering to
the following
final states $NR, RR, \Delta\Delta\pi, NN\rho$ and $NN\omega$ where
$R$ denotes baryon resonances $\Delta, N^*(1440)$ or $N^*(1535)$.
As one would expect from the decomposition
of the total inelastic cross sections
discussed in the previous section, the rate of the quasi-3$\pi$ channel
is the highest at the very beginning of the reaction
when $t\leq 1$ fm/c,
but is closely followed by the quasi-2$\pi$ and quasi-1$\pi$ channels.
Since particles loses their energies after several collisions the single
baryon resonance production soon becomes dominant. The collision rates
for producing $\rho$ and $\omega$ are, however, about an order of magnitude
smaller than the quasi-4$\pi$ channel. Most collisions cease after about
10 fm/c except the single-resonance production channel
$NN\rightarrow NR$ which lasts a little longer.
For comparison, it is worth noting that the maximum reaction rate reached
in the collision
is about an order of magnitude higher than that in heavy-ion
collisions at Bevalac and SIS/GSI energies\cite{li91a,dani95}.

The rescattering or absorption of produced baryon resonances
and mesons are very important in heavy-ion reactions at AGS energies.
In Fig.\ 13 we show the reaction rates for nucleon+resonance,
resonance+resonance and meson+meson scattering.
It is seen that the nucleon+resonance scattering  start earlier and dominate
during most of the collision time. The meson+meson and
resonance+resonance rescattering start later as it
takes some time to create these
secondary particles. After about 10 fm/c the nucleon+resonance
and resonance+resonance scattering are almost over. One also notices that
meson+meson scattering last longer mainly due to the large $\pi+\pi$
elastic cross section, although the rate decreases towards the later time
of the collision. All these features are
what one would expect and easily understandable.
The relatively low rate for baryon resonance+baryon resonance
collisions needs, however, some explanations.
The most important reason is that the final states in
resonance+resonance collisions are limited only to $NN$, $NR'$ and $K^+X$
in the present version of the model, instead of allowing for
the production of up to 4 quasipions as in nucleon+nucleon collisions.
In addition, due to the large spin degeneracy in the initial state,
the cross sections obtained from the detailed balance relation
for the $RR\rightarrow NN(R')$ reactions are relatively
small compared to their reverse reactions.

In Fig.\ 14, we show the decay rate of baryon resonances together with
their formation rate from $\pi+N$ collisions. An interesting
feature of these rates is that they last a longer time than
hadron+hadron collisions due to the large meson+baryon cross sections.
Another feature is that the small difference between the rates of
decay and formation of baryon
resonances makes the apparent lifetimes of these
resonances much longer than their lifetimes in free space.
This is rather clear from
the evolution of the pion and baryon resonance multiplicities
shown in Fig.\ 15. It is also seen that about half of the baryons
are in their
excited states at about 4 fm/c, which is the instant of the highest
compression as we will discuss in the next section.
The abundance of baryon resonances is dominated by
$\Delta(1232)$ resonances as one would expect in this energy range.
However, it should be stressed that the population of $N^*(1440)$ and
$N^{*}(1535)$ resonances is also significant. These higher resonances
serve as an energy reservoir and thus play an important role in the production
of particles, especially those having higher energy thresholds\cite{li94}.
It is also interesting to note that the number of pion-like particles
(free pions + baryon and meson resonances) is about the same
as the initial number of resonances at the time of maximum compression.

In Fig. 16, we show the time integrated total number of
hadron-hadron collisions as a function of the center-of-mass
energy $\sqrt{s}$ of the colliding pairs
for $N+N, N+R, R+R, M+B$ and $MM$ collisions, where
the meson+baryon ($M+B$)
and meson+meson ($M+M$) collisions
include contributions from both baryon and
meson resonances. It is seen that $N+N$
collisions start at about $\sqrt{s}=1.8$ GeV and peak at
about $\sqrt{s}=2.2$ GeV. The peak is mainly due
to the large number of elastic scattering. The position of the peak shifts to
higher energies and its hight decreases substantially in
$N+R$ and $R+R$ reactions. The reasons are mainly the following. First of all,
we have assumed that the elastic
cross sections for $N+R$ and $R+R$ collisions are the same as
for $N+N$ collisions at the same center-of-mass energy $\sqrt{s}$.
Since the elastic cross section decreases quickly as the center-of-mass energy
increases, the heavier masses of the baryon resonances
then result in smaller elastic cross sections for
$N+R$ and $R+R$ collisions.
In addition, as we have mentioned earlier, the number of final states in
$N+R$ and $R+R$ collisions are reduced compared to that in $N+N$ collisions.
For meson+baryon reactions, the collision number
distribution have
two distinct peaks at $\sqrt{s}\approx 1.2$ GeV and $\sqrt{s}\approx 2$ GeV,
respectively. The first peak apparently corresponds to the formation
of the $\Delta(1232)$ resonance, while the second peak has contributions also
from the $N^*$ resonance and direct $\pi-N$ collisions. The meson+meson
collision number has a peak at about the $\rho$ mass
$\sqrt{s}\approx m_{\rho}=$0.77 GeV.

The above examination on the collision dynamics indicates that the model
is working well under control, regardless of the question whether some
of the assumptions made in the model can be improved.
This gives us further confidence about the model.

\subsection{Formation of superdense hadronic matter}
Depending on the nature of the quark-gluon plasma phase transition at high
baryon densities, the crossover energy between the hadronic and quark-gluon
phases may occur anywhere between $E_{lab}/A=2-10$ {\rm
GeV}\cite{rischke,glen}.
It was pointed out recently that the current Au+Au reactions at 11.6 GeV/c
may have already overshot the transition region\cite{gyu}. It is therefore
of great interest to critically examine the baryon, meson and
energy densities reached in these collisions. In this section, we
study the local baryon, meson and energy densities in head-on collisions
of Au+Au at $p_{beam}/A=$11.6 GeV. It will be shown that
in these collisions a superdense
hadronic matter with a density higher than the
critical density for the phase transition can be formed in a sufficiently
large volume and for a sufficiently long time.
We will also study the dependence of the size and lifetime of the
superdense hadronic matter on the nuclear mean field.

The standard test particle method\cite{wong82}
is used to calculate the global density $\rho_g$ in the
nucleus-nucleus c.m. frame on a lattice of size $40-40-48$
with a volume of 1 ${\rm fm}^3$ for each cell.
In the test particle method, one replaces the continuous
phase space distribution function with a finite number of test
particles representing individual phase space cells, i.e.
\begin{equation}\label{test}
f(\bold{r},\bold{p},t)\simeq \frac{1}{N_{t}}\sum_{i}\delta(\bold{r}
-\bold{r}_{i}(t))\delta(\bold{p}-\bold{p}_{i}(t)),
\end{equation}
where $\bold{r}_{i}$ and $\bold{p}_{i}$ are the coordinates and momenta
of the test particles. $N_{t}$ is the number of test
particles per nucleon,
we usually use $N_t=100$ for Au+Au collisions.
The global baryon, meson and energy density function
$\rho_g^{b}$, $\rho_g^m$ and $e_g$ are
evaluated via
\begin{eqnarray}
&&\rho_g^b(r,t)=\sum_{i=N,\Lambda,\Sigma}
\int \frac{d^{3}\bold{p}}{(2\pi)^{3}}f_i(\bold{r},\bold{p},t)
+\sum_{i=\Delta, N^* }\int \frac{d^{3}\bold{p}}{(2\pi)^{3}}\frac{dm}{2\pi}
f_i(\bold{r},\bold{p},t),\\
&&\rho_g^m(r,t)=\sum_{i=\pi,K,\eta,\omega}
\int \frac{d^{3}\bold{p}}{(2\pi)^{3}}f_i(\bold{r},\bold{p},t)
+\int \frac{d^{3}\bold{p}}{(2\pi)^{3}}\frac{dm}{2\pi}
f_{\rho}(\bold{r},\bold{p},t),\\
&&e_g(r,t)=\sum_{i}\int \frac{d^{3}\bold{p}}{(2\pi)^{3}}(p^2+m^2)^{1/2}
f_i(\bold{r},\bold{p},t)+\sum_{j}\int \frac{d^{3}\bold{p}}{(2\pi)^{3}}
\frac{dm}{2\pi}(p^2+m^2)^{1/2}
f_j(\bold{r},\bold{p},t).
\end{eqnarray}
In the last equation, $i$ runs over all particles with fixed masses
while $j$ runs over all particles with variable masses.

The local baryon, meson and energy densities
in each cell are then obtained
from $\rho_{l}^b=\rho_{g}^b/\gamma$,
$\rho_{l}^m=\rho_{g}^m/\gamma$,
and $e_{l}=e_{g}/\gamma$,
where $\gamma$ is the Lorentz factor of each cell,
\begin{equation}
\gamma=\sqrt{(\sum_{i}E_i)^2-(\sum_i\vec{p_i})^2}/\sum_iE_i.
\end{equation}
In the above, $i$ runs over all particles in the cell.
For most parts of the colliding nuclei, the use of local
densities can eliminate the trivial compression effects
due to the initial Lorentz contraction of the two nuclei.
However, in the most central cell where the
two nuclei are in touch and streaming towards each other,
the $\gamma$ factor is almost 1.
Therefore, we have at the very beginning of the collision,
$\rho_l^b\approx \rho_g^b=2\gamma_0\rho_0$
where $\gamma_0$ is the center of mass Lorentz factor of the
colliding nuclei, and for Au+Au at $P_{beam}=11.6 $ GeV/c, one has
$\gamma_0=2.6$. Fortunately, this initial high baryon density is only
in a very small region at the very beginning and is soon destroyed by
the violent collisions that follow. In the cascade model calculations,
the initial high baryon density in the most central cell is simply a matter of
presentation. For calculations including the density-dependent mean
field we have found from comparisons with
calculations using the scalar density that the initial high baryon density
has no effects on the collision dynamics and the final observables.

We show in Fig.\ 17 the evolution of the
local baryon density distribution in the reaction plane
using the cascade mode of ART 1.0 for the head-on
collision of Au+Au at $P_{beam}/A$=11.6 GeV/c.
The outmost contour is for $\rho_l=0.5\rho_0$, and the numbers in the figure
are the densities of the corresponding contours in units of $\rho_0$.
The two Lorentz contracted nuclei are set in touch at t=0. It is seen that
the initial high density region bounded by the contour with
$\rho=2\rho_0$ is very small. The two nuclei
soon form a compressed disk  and reach the maximum compression
of about $9\rho_0$ at about 4 fm/c.
The compressed hadronic matter then starts to expand at about 6 fm/c.
We see that up to 10 fm/c the longitudinal expansion is significantly
stronger than the transverse one.
This observation seems to indicate that the isotropically expanding,
spherical fireball model widely used for
describing heavy-ion collisions at AGS
energies may need to be improved. It should be mentioned that the
main features observed here are in general
agreement with those extracted from the ARC calculations\cite{arc}.

The local densities are well defined and are the relevant quantities
for discussing the interesting physics of the phase transition
to a quark-gluon plasma. However, the current model is still not
fully relativistically covariant. To estimate the minimum compression
reached during the collision for a given system at a fixed beam energy and
impact parameter, we show in Fig.\ 18 the scalar baryon density
$\rho^b_s$ defined as
\begin{equation}
\rho_s^b(r,t)=\sum_{i=N,\Lambda,\Sigma}
\int \frac{d^{3}\bold{p}}{(2\pi)^{3}}f_i(\bold{r},\bold{p},t)\frac{m}{e}
+\sum_{i=\Delta, N^* }\int \frac{d^{3}\bold{p}}{(2\pi)^{3}}\frac{dm}{2\pi}
f_i(\bold{r},\bold{p},t)\frac{m}{e},
\end{equation}
where $e$ is the energy of a baryon. In the limit that the size
of each cell is very small such that it can
only accommodate one particle the local density becomes equal to the
scalar density. It is seen that the main features in the evolution
of the scalar density are similar to those of the local density.
It is interesting to note that the initial high density region around the
most central cell in the local density plot does not appear here as
one would expect. We have
found that the maximum scalar density reached at 4 fm/c is
about $6.5\rho_0$ which is significantly smaller than the maximum
local density reached in the collision.

The local density of mesons (mostly pions) can be as high as that of nucleons
during the collision. The evolution of the meson density distribution
in the
reaction plane is shown in Fig.\ 19. Mesons mainly populate the
central region for $t\leq 2$ fm/c, and they spread out quickly soon after that
and finally occupy the whole collision volume.
The maximum meson density of about $4.5\rho_0$ is reached at about 4 fm/c.
As we have discussed earlier, at this time about half of
final pions are still bound in resonances.
It is also interesting to see that mesons start to expand earlier and
faster than baryons.
After about 8 fm/c, the meson sector of the system has already been
expanding almost isotropically. In principle, the geometry and time scale of
the expanding hadronic matter studied here can be compared directly
with the interferometry measurements to further test the model.

To form a quark-gluon plasma, it is necessary to achieve a high
local energy density in a sufficiently large volume and
for a sufficiently long time such that the
initial plasma droplets can grow.
The current estimate for the critical baryon and
energy densities at which the QGP may form are about
$5\rho_0$ and 2.5 {\rm GeV/fm}$^3$, respectively\cite{wong}.
{}From the above studies on the baryon and meson densities, we have
seen that the superdense hadronic matter
occupies a large volume and for a rather long time.
As to the possible phase transition to the quark-gluon plasma,
more complete and quantitative information about the local energy density
of the superdense hadronic matter are needed.
Fig.\ 20 shows the evolution of the local energy density in the reaction
plane for the head-on collision of Au+Au at $p_{beam}=11.6 $ GeV/c.
The numbers in the figure are the energy densities in units of GeV/fm$^3$.
It is seen that a small, thin disk of energy density higher than
2.5 GeV/fm$^3$ has already been formed at about 2.0 fm/c.
It grows until about 4 fm/c, then starts
to decay, and at about 7 fm/c it disappears due to expansions.
The superdense hadronic matter
has thus a lifetime of about 5 fm/c and a maximum volume
of about 200 fm$^{3}$ at about 4 fm/c.

The results obtained from the cascade calculations
seem to indicate that in head-on collisions of
Au+Au at $p_{beam}/A$=11.6 GeV/c, the conditions for forming a quark-gluon
plasma has indeed been reached.
However, this also marks the breakdown of the
hadronic models and the need to
include new degrees of freedom.  It is thus a great challenge to actually
convert this superdense hadronic matter into the quark-gluon plasma.
Without introducing explicitly the quark and gluon degrees of freedom
in the model make this impossible.
Nevertheless, before one can study the properties of the
quark-gluon plasma, it is necessary to investigate how and to what degree the
properties of the superdense hadronic matter may be affected by the
nuclear mean field, which is an important piece of physics neglected
in cascade calculations. In the following, we shall compare
results obtained with and without the mean field.

First of all, we have found that the general features in the evolution
of the baryon, meson and energy density distributions are very similar
in the two calculations. However, important quantitative
differences exist. Fig.\ 21 shows the evolution of the
local baryon and energy densities in the most central cell during
the collision of Au+Au at $P_{beam}/A$=11.6 GeV/c and b=0.
Results from the cascade calculations
are compared with those from calculations
using the soft nuclear equation of state with $K=200$ MeV and the
stiff one with $K=380 $ MeV. As expected, the mean field
has almost no effect in the early
stage of the collision when the kinetic
energy is much higher than the potential energy.
Significant differences appear soon after about 2 fm/c.
In the cascade case, it is seen that a maximum baryon density
of about $9\rho_0$ and a maximum
energy density of about 3.6 {\rm GeV/fm}$^{3}$ are reached at
about 4 {\rm fm}/c. The matter in the high energy density region
lasts for about 5 {\rm fm}/c. The
volume of the high density region are, however,
significantly reduced by the repulsive mean field. Furthermore,
the lifetime, maximum baryon and energy densities are
reduced to about 3 fm/c, $7\rho_0$ and 2.6 {\rm GeV/fm}$^{3}$, respectively,
when using the soft nuclear equation of state.
With the stiff equation of state the reduction is even larger.
Since the stiff equation of state violates causality already at
about $3\rho_0$, we will only use the soft equation of state in the following.
The reduction of the maximum baryon and energy densities due to the
mean field may be large enough to affect significantly
the collision dynamics.

Additional information about the importance of the nuclear mean field
can be obtained by comparing the fraction of particles with local energy
densities $e_l\geq 2.5 $ GeV/fm$^3$ from calculations with and without
the mean field. This is shown in Fig.\ 22. In the cascade calculation
about 60\% of particles are in the high energy density region
at the instant of the maximum compression. The soft
equation of state reduces it to about 30\%, and this reduction
persists during the expansion phase of the collision.

\subsection{Thermalization and radial flow }
To what degree the superdense hadronic system
is thermalized and what is the time scale for achieving
thermal equilibrium are among the most important questions
in heavy-ion collisions at AGS. Answers
to these questions are useful both for
developing static models and for interpreting the experimental data.
For example, in thermodynamic models one usually assumes that
there exists a thermalized freeze-out stage when particles in
the expanded system cease to strongly interact with each other.
In view of the apparent success of the thermodynamical
approach (e.g.\ \cite{munz}) in explaining a large amount of data
from heavy-ion collisions
at AGS energies we study in this section the question of
thermalization. In addition, the large thermal and compressional pressure
in the superdense hadronic matter is found to induce
a radial flow during the expansion phase. This is already qualitatively
clear from our discussions in the previous section, and we
study here more quantitatively the radial flow.

We consider all hadrons in a sphere of radius 2 fm around the origin in the
center of mass frame for head-on collisions of Au+Au at $p_{beam}/A=11.6$
GeV/c. The degree of thermalization can
be measured by the ratio
\begin{equation}
R=\frac{\sum_{i}(p_{ix}^2+p_{iy}^2)}{2\sum_{i}p_{iz}^2}.
\end{equation}
Then $R=1$ is a necessary, although not a sufficient, condition for
thermal equilibrium. It is also a measure of the stopping power.
In the bottom window of Fig.\ 23, we show
this ratio as a function of time. For all three calculations
based on the cascade, the soft and the stiff equation of state,
the ratio reaches one at
about 10 fm/c. From our early discussions on the reaction rates for various
channels, it is clear that this is about the time when
most hadronic collisions
have stopped. It is therefore reasonable to use the concept of
thermal equilibrium at the time of freeze-out.

It is interesting to note that at about 4 fm/c when the
system reaches its maximum compression the ratio R is
about 0.7, indicating that the superdense hadronic
system has not reached thermal equilibrium.
To estimate the temperature of particles in the above sphere at the time of
thermalization, we show in the middle window the
scaled average kinetic energy per particle $<\frac{2}{3}E_{k}>$.
At freeze-out, this quantity can give
a good estimate of the temperature. Before freeze-out,
the evolution of this quantity reflects the collision dynamics.
It is seen that the kinetic energy decreases quickly in the compression phase
due to the creation of baryon resonances and other particles as well as
various scattering. In the expansion phase, high energy
particles near the surface of the sphere move out continuously
due to the radial flow. The average
kinetic energy in the sphere thus gradually decreases in the expanse phase.
At the time of freeze-out this quantity is about 120 MeV and slightly depends
on the model. This value is very close to the temperature
extracted from the inverse slope of the measured
particle spectra. It is interesting to note that both the time
scale for thermalization and the temperature at freeze-out are not very
sensitive to the equation of state. However, the temperature from
the cascade is about 30 MeV higher than that from calculations
with the mean field due to the neglect of the potential energy in the former.

In our calculations, the system
has undergone sufficient expansion before reaching thermal
equilibrium  at
freeze-out, and this is in agreement with the conclusion
reached from studies of the AGS data based on
the thermodynamical model \cite{munz}.
More quantitative information about the expansion can be seen from the
upper window where the number of particles in the central sphere
is plotted as a function of time. General features of this plot
are the same as that of the central density plot discussed earlier. It is seen
that the number of particles in the sphere at freeze-out is
only about 30\% of that at the maximum compression indicating a strong
expansion before the freeze-out.

The velocity profile of the radial flow at freeze-out can be
expressed by
\begin{equation}\label{fl}
\beta(r)=\frac{1}{N_r}\sum_{i=1}^{N_r}\frac{\vec{p_i}}{e_i}\cdot
\frac{\vec{r_i}}{r_i},
\end{equation}
where $N_r$ is the number of particles in a spherical
shell of size $\Delta r=0.5$ fm at
a radius $r$. The momentum $\vec{p}_i$ of each particle
has a flow component and a random thermal component
\begin{equation}
\vec{p}_i=\vec{p}_{if}+\vec{p}_{it}.
\end{equation}
In the limit that there are a large number
of particles in each shell, the summation in
Eq.\ (\ref{fl}) cancels out the thermal part, i.e.
\begin{equation}
\sum_{i=1}^{N_r=\infty}\frac{\vec{p}_{it}}{e_i}\cdot \frac{\vec{r}_i}
{r_i}=0,
\end{equation}
and Eq.\ (\ref{fl}) reduces thus to the flow velocity $\beta_f(r)$,
\begin{equation}
\beta(r)\approx\beta_f(r)=
\sum_{i=1}^{N_r=\infty}\frac{\vec{p}_{if}}{e_i}\cdot \frac{\vec{r}_i}
{r_i}.
\end{equation}
Since we usually use a large number of test particles,
Eq.\ (\ref{fl}) gives a good estimate of the radial flow profile.
In Fig.\ 24, we show the estimated flow
profile and the corresponding radial density distributions
for baryons and pions.
The left window is for the cascade and the right one is for calculations
using the soft equation of state.
The radial flow velocity is proportional to the radius,
but there seem to be two distinct
slopes at radii smaller and larger than about 6 fm. Another
interesting feature is that pions seem to flow a little faster than baryons
due to their lighter masses.
It is also seen that the difference between the flow profiles from
calculations with and without the mean field is rather small,
which may indicate that the thermal pressure dominates at freeze-out.
As seen from the upper window,
more particles are at larger radii due to the repulsive
mean field, so more energy are
stored in the radial flow in the case with the mean field.

For comparisons, we mention here that results of
a recent hydrodynamical model analysis of a large amount of data from
central collisions of Si+Au at $p_{beam}/A=14.6$ GeV/c are consistent
with the above discussions. Assuming a freeze-out radius of about 7 fm,
a freeze-out temperature of 120 (140) MeV and a linear flow velocity
profile, the extracted maximum flow velocity is
about 0.58 (0.5) c\cite{munz}, which agrees with
the flow profile we have extracted.

\subsection{Stopping power and inclusive observables}
The formation of a quark-gluon plasma or a mixed phase of hadrons, quarks and
gluons is expected to reduce the
pressure of the system and leads to a softened
equation of state. It is therefore interesting to search for
experimental observables that are sensitive to
the nuclear equation of state. In this section, we study
the sensitivity of inclusive, single particle observables to the
equation of state by comparing calculations with and without the mean field.
However, we should mention again that
present models for heavy-ion collisions at AGS energies are probably
inconsistent.
Namely, we have shown in the above that the conditions for
forming the quark-gluon plasma may have been fulfilled in central
collisions at AGS energies. Without actually including the phase
transition to the quark-gluon plasma and the hadronization back
to the hadronic phase in the model, we are limited to discuss only the
hadronic physics.

Specifically, we shall study the rapidity and transverse momentum distributions
of protons and pions. We leave the study of kaon distributions to the
next section in central collisions of Au+Au
at $p_{beam}/A=$11.6 GeV/c. For references, we also compare with
preliminary data from the E802/E866 collaboration\cite{e866}.
These data were taken using the kinetic energy in the zero
degree calorimeter as a trigger on the impact parameter.
The data for central collisions roughly correspond to the upper 4\% of the
total cross section or impact parameters smaller than about 3 fm. Since the
data still have a 15\% systematic errors besides the statistical ones,
we have not let the generated events going through the zero
degree calorimeter to simulate the experimental trigger conditions.
Instead, we perform calculations with impact parameters smaller than 3 fm
and compare the data with calculations using the impact parameter of 2 fm
which is about the average impact parameter of the central collisions.
This is probably sufficient for our current purposes.

First, let us study the proton transverse momentum distribution.
In Fig.\ 25, we show the calculated spectrum by the round symbols in
the left window using the cascade and the right window using the soft equation
of state. The least square fit to the
data of ref.\ \cite{e866} are shown by the solid lines.
In the cascade case, the
calculation agrees with the data rather well except in the low transverse
momentum region. The larger theoretical values are probably
due to the fact that we count all protons in the final state although some of
them are still bound in light clusters, while the data are for protons only.
To estimate the number of bound protons, we apply a
cut to the local baryon density such that only protons
with $\rho_l< 0.2 \rho_0$ are counted. The
spectra calculated in this way are shown with the cross symbols which
agree well with the data even at low transverse momenta.
The spectra calculated with
the soft equation of state show an increase of about 25\% mainly
at large transverse momenta compared with that from the cascade
as a result of the repulsive mean field at high densities.
This difference, however, is comparable to the combined
statistical and systematical errors of
the data. Nevertheless, it is promising that the proton spectra at high
transverse momenta may be useful for studying the equation of state at the
level of sensitivity of about 25\%. As we will show later, a much
clearer sensitivity at the level of 250\% can be obtained
from an event by event analysis of the baryon transverse collective flow.

In Fig.\ 26, we show the calculated proton rapidity
distributions. As one would expect, the mean field leads to less stopping.
This observation is consistent with our early discussions about the
mean field effects on the creation of superdense hadronic matter.
It is seen that without applying the local density
cut at $0.2\rho_0$, the mean field reduces the height but increases
the width of dN/dy at midrapidity by about 15\%.
With the local density cut, the
calculation with the mean field can better reproduce the data, but again
the difference between the two calculations is comparable to
current systematic errors in the data. As we have stated earlier, the
above calculations are carried out at a fixed impact parameter of 2 fm.
To quantify the impact parameter dependence of the stopping power in central
collisions, we show in Fig.\ 27 the proton rapidity distributions at impact
parameters of 1 fm and 3 fm. In both calculations with and without
the mean field, the height of the rapidity distribution
at the midrapidity increases by about 25\% as the impact
parameter changes from 3 fm to 1 fm.
With an impact parameter of 3 fm, a clear bump due to spectators at
the projectile and target rapidities can be seen. The variation of the
rapidity distribution at these two impact parameters caused by the mean
field is again at the level of 15\%.

Pions are copiously produced in heavy-ion collisions
at AGS energies and are expected
to play an important role in the stopping and thermal equilibration
of the system.
The study of pion rapidity and transverse momentum spectra
provides thus useful information about the collision
dynamics, the properties of the superdense hadronic matter, and
the particle production mechanisms.
In Fig.\ 28 and Fig.\ 29, we compare calculated $\pi^+$ and $\pi^-$
transverse momentum spectra using the cascade and soft equation of
state with each other and also
with the least square fit to the data (solid lines).
The rapidity range for each spectrum is indicated in the figure.
It is seen that both calculations
can reproduce the data reasonably well except at rapidities far
from the center of mass rapidity,
which is about 1.72 for the Au+Au reaction at $p_{beam}/A$=11.6 GeV/c.
Again, the variation between calculations
using the cascade and the soft equation of state is on the order of 20\%
which is comparable to current experimental errors. The statistics
in the calculations can be significantly improved by mixing events from
different impact parameters. But main features of the spectra
are expected to remain unchanged.
We will leave a more detailed study on pion spectra,
including the contribution of resonance decays at freeze-out and
the Coulomb effects on the difference between $\pi^+$ and $\pi^-$ spectra to
a separate paper.

In Fig.\ 30, we compare the rapidity distributions of $\pi^+$ calculated
at an impact parameter of 2 fm using the cascade and the soft equation of
state with the experimental data. The calculation with the soft equation
of state seems to be closer to the data, but there exists some
discrepancies. It is worth mentioning that
the experimental rapidity distribution was obtained by first
fitting the transverse
momentum spectra with an exponential function or a sum of two exponential
functions and then extrapolating and integrating over the whole rapidity range.
While in our calculations we simply count pions in each rapidity bin
without any possible bias from the extrapolation process. It is seen that
the difference between calculations with and without the mean field is
again on the order of 20\%, which is rather similar to the situation at
the Bevalac or SIS/GSI energies\cite{dani91,li93b}.
The impact parameter dependence of the pion
rapidity distribution is shown in Fig.\ 31. The dependence here is rather
weak compared to that of the proton rapidity distribution. This is mainly
due to the fact that there is an almost complete overlapping between the
target and the project in collisions with impact parameters smaller than 3 fm
and the large number of pions produced in these collisions.

\subsection{Mechanisms for $K^+$ production}

Mechanisms for kaon production at AGS energies
have been a subject of intensive studies in the past few years.
In particular, the enhancement of $K^+/\pi^+$ ratio in heavy-ion
collisions with respect to $pp$ and $pA$ reactions has been
attracting much attention due to the possible role
of strangeness as a signature of the quark-gluon plasma\cite{raf,koxia}.
However, no definite conclusion has been reached
due to the complexity
of the collision dynamics at AGS energies and our incomplete knowledge
about the properties of hadrons and their interactions
in hot dense matter.

In Fig.\ 32, we show the contributions to
$K^+$ production from various collision
channels as a function of the center-of-mass energy of the colliding pairs
in Au+Au collisions at $P_{beam}/A=11.6$ GeV/c.
It is seen that large contributions come from meson+baryon
and meson+meson collisions, besides that
from baryon+baryon collisions.
This is in agreement with the predictions from ref.\ \cite{ko} based on
the hydrochemical model calculations. More quantitatively,
about 35\%, 40\% and 25\% of the produced
kaons come from baryon+baryon, meson+baryon
and meson+meson collisions, respectively. The relative importance of different
$K^+$ sources seen here does not agree completely with
results from either the RQMD or the ARC.
Since there are many assumptions in each model, it is not apparent
to us at this time what is the cause for the difference
among these models. It is certainly of great interest to resolve this
difference in future studies. However, this can only
be possible when all assumptions used in all models are made clear.

For ease of future comparisons with other models and also for studying the
kaon production dynamics, we show in Fig.\ 33 the
$K^+$ production rate as a function of time for various channels.
The calculation is carried out for head-on collisions at $p_{beam}/A=11.6$
GeV/c using the cascade. It is seen that at $t\approx 1.0$ fm/c the
production rates due to nucleon+nucleon, meson+baryon (including resonances)
and nucleon+baryon resonance are almost the same.
Soon after that the $K^+$
production rate is dominated by meson+baryon collisions which also last
a longer time. The $K^+$ production rates due to
nucleon+baryon resonance, nucleon+nucleon and
baryon resonance+baryon resonance decrease to zero after about 6 fm/c.
Meson+meson collisions come in later, but they contribute significantly
to the $K^+$ production especially during the expansion phase
of the collision. The time dependence of $K^+$ production studied here
may be useful for further studies on the production mechanisms for
kaons. Moreover,
kaons produced at different times through different channels may also
contribute to different parts of the kaon spectra. We, however, leave
the more detailed discussions on this subject to future studies.

We now go back to the question of comparing kaon spectra calculated
with and without the mean field with the preliminary data from the E802/E866
collaboration. Fig.\ 34 shows the comparison on the kaon transverse
momentum distributions. The solid lines in both windows are the
exponential fit to the data to guide the eye. It is seen that both calculations
agree with the data reasonably well. However, there are
some discrepancies between calculations and experimental data
especially at high transverse momenta. Within error bars of the
calculations the two calculated spectra are almost indistinguishable.
Fig.\ 35 shows the $K^+$ rapidity distributions. The data
at rapidities larger than 1.72 are the reflections of those at rapidities less
than 1.72. Except at rapidities near the center-of-mass rapidity the two
calculations are very close to each other. Even at the midrapidity the
sensitivity to the equation of state is only about 15\%.
This situation is rather different from that
at SIS/GSI energies where subthreshold kaon production is very
sensitive to the nuclear equation of state\cite{aich,huang,fang,mar,likaon}.
Since kaon production at AGS energies is high above the threshold,
the sensitivity of kaons to the equation of state is not expected to be
very different from that of nucleons and pions.
To be complete, we show the impact parameter dependence of kaon production
in Fig.\ 36. The sensitivity of kaon multiplicity to the impact
parameter is at the level of 20\% in central collisions, which is
rather similar to that of pions.  The same level of sensitivity
to both the impact parameter and the equation of state in central collisions
makes it even harder to learn about the equation of state from
these observables. In contrast to the inclusive
observables discussed above,
collective observables are very sensitive to the equation of state
as we shall show in the next section.

\subsection{Baryon transverse flow}
Hinted by the findings at Bevelac energies
that the collective variables or correlation functions, unlike single
particle observables, are very sensitive to the nuclear equation of state,
we now turn to the analysis of the baryon transverse collective
flow\cite{dani}.
Here we use the standard flow analysis in transport models
as in ref.\ \cite{bertsch}. It has been shown recently
in ref.\ \cite{gale} that
an improved flow analysis with the explicit conservation of
reaction plane and angular momentum in an individual hadron-hadron
collision results in an increase of the transverse momentum by about
8\% to 23\% at Bevalac energies. This small enhancement
is not expected to affect our discussions and conclusions.

First, we show in Fig.\ 37 the average transverse momentum of
nucleons in the reaction plane as a function of rapidity for the
collision of Au+Au at $P_{beam}/A=$11.6 GeV/c and at impact parameters
of 2, 6 and 10 {\rm fm}.
It is seen that the collective flow is the strongest in midcentral
collisions in both calculations.
Most importantly, significant differences exist between
calculations with and without the mean field for collisions at
all three impact parameters. In particular, the flow
parameter defined as
\begin{equation}
F=(\frac{dp_x}{dy})_{y_{cm}}
\end{equation}
is about a factor of 2.5 larger
in the case with the mean field. The strength of
the so-called ``bounce-off'' effect at target or projectile
rapidities is also much stronger in calculations with the mean field.
To measure the strength of the ``bounce-off'' effect, we define
the average total in-plane transverse momentum as
\begin{equation}
< P_x >=\int_{-2}^2 |dp_x/dy|dy.
\end{equation}

To see how the collective flow is a sensitive probe of the reaction
dynamics in the high density region, we show in Fig.\ 38 and Fig.\ 39
the flow parameter $F$ and the total in-plane transverse momentum $< P_x >$
as functions of time for Au+Au reactions
at impact parameters of 2 {\rm fm} and {\rm 6 fm}, respectively.
For comparisons, the central densities are also shown in the lower windows
of Fig.\ 38 and Fig.\ 39. It is clearly seen that the flow is mainly
generated in the high density region and does not change
in the expansion phase. The increase of the
total in-plane transverse momentum
during the expansion phase is much more substantial in the case
with the mean field than in the cascade case. This indicates
that the mean field effect is large in the expansion phase as we
have expected. The total in-plane transverse
momentum reaches its final value
after about 15 fm/c. The ratio of final flow parameters
in the calculations with and without the mean field is
about 2.5 at both impact parameters.
This difference is large enough for distinguishing the predictions
from different
models. For collision at an impact parameter of 6 fm, the flow parameter
decreases slightly before reaching its final value.  This is mainly
due to the reflection of hot baryons from the cold spectator
nucleons in collisions at large impact parameters.

It is interesting to note that a clear signature of transverse collective
flow has been discovered recently in Au+Au reactions at $p_{beam}/A$=
11.4 GeV/c by the E877 collaboration\cite{E877}. In this experiment the
distribution of the normalized transverse energy dipole moments is found
to shift systematically toward finite values for more central collisions.
We are planning to make a detailed comparison with these data in the near
future. Furthermore, the preliminary data of the E877
collaboration indicates a smaller transverse momentum in the reaction
plane than the predictions from both present and other models
\cite{E877-95,zhang}, which may
indeed indicate that the equation of state is significantly
softened due to the formation of the quark-gluon plasma.

\section{Summary}
In summary, we have developed a new relativistic transport model (ART 1.0)
for heavy-ion collisions at the AGS energies.
This model is an extension of the very successful BUU model for intermediate
energy heavy-ion collisions by including more baryon and meson resonances
as well as interactions among them. Comparing to the well-known
ARC model \cite{arc},
the present model has more resonances which are found
to be important for the collision dynamics and particle production
at AGS energies. Moreover, we have used consistently finite mass
distributions for these resonances which have significant effects on
particle spectra. In addition, a selfconsistent mean field for baryons is
included. However, comparing to another well-known RQMD model \cite{rqmd},
where all resonances up to mass of 2.0 GeV are explicitly included, we
do not have as many baryon resonances.  Nevertheless, we have
partially included the effects of these resonances by using
meson+baryon cross sections calculated from
the implicit formation of higher resonances.
As we are mainly interested in developing a reliable model for energies
up to about 15 GeV/nucleon, we do not invoke the string/rope mechanisms
for particle production as in RQMD.

Due to the lack of some hadron-hadron collision data, the input to the model
still has some uncertainties. Several assumptions used in the model may
need to be improved in future versions of the model.
In particular, interactions among resonances are largely unknown and have been
assumed to be the same as nucleon-nucleon interactions.
Additional hadrons, such as $\bar{p},~ \bar{\Lambda}$, etc, are
to be included in the model. At AGS energies the resonance population
is as large as that of nucleons in the compression phase,
the assumption of the same mean field for nucleons and resonances
may need to be improved. Currently, the finite formation time for
produced particles has only been partially taken into account through
the finite lifetime of the resonances. A more consistent
method may be necessary. Also, it is of interest to include the mean field
potential for pions\cite{xiong,koch} and kaons\cite{gqli} as well as the Bose
enhancement factor for pions\cite{gerd}.
The most inconsistent, but also probably the most interesting
feature shared by this model and others is
the prediction that the conditions for forming the quark-gluon plasma
have already been reached in heavy-ion collisions
at AGS energies. In particular, at the maximum
baryon density of about
$9\rho_0$ and meson density of $4.5\rho_0$ hadrons are squeezed on top of
each other and probably melt into the quark-gluon plasma. At this high
density hadronic models naturally breaks down. It remains a
great challenge to actually include the quark and gluon degrees of
freedom into the model and to study the phase transition.

With the above reservations about the validity of the model, we have
studied several aspects and issues of heavy-ion reactions
at AGS energies. In particular, we have explored
effects of the nuclear mean field on the collision dynamics, the formation
of superdense hadronic matter, single particle observables as well as
the radial and transverse collective flow.
It is found that the mean field significantly reduces the
maximum energy and baryon densities of the superdense hadronic matter formed
in the collision. Inclusive single particle observables are sensitive to the
equation of state only at the level of 20\%.
On the other hand, the transverse collective flow
of baryons are found to be very sensitive to the equation of state.
In particular, including a soft nuclear equation of state increases
the flow parameter by a factor of 2.5
which is large enough to distinguish the predictions from different
theoretical models.
Moreover, this sensitivity also indicates that the collective flow
can be a useful probe of the formation of the
quark-gluon plasma at AGS energies. Preliminary data have shown
a smaller baryon transverse flow in the reaction plane than model
predictions, and this may indeed indicate the significant
softening of the equation of state
due to the formation of the quark-gluon plasma.
It is very encouraging that the present model has been very
useful for studying the interesting physics in high energy
heavy-ion collisions.
\medskip

\section{Acknowledgement}
We would like to thank many participants at the program
``Hot and dense nuclear matter'' held at the National Institute for Nuclear
Theory in Seattle for interesting discussions. In particular, we
are indebted to W. Bauer, L.P. Csernai, P. Danielewicz,
S. Das Gupta, V. Koch, G.Q. Li, J. Randrup, D. Strottman, and Gy. Wolf
for their constructive suggestions and criticisms.
This work was supported in part by NSF Grant No. PHY-9212209
and the Welch Foundation Grant No. A-1110.

\begin{table}
\caption{Isospin cross section parameters}
\label{isospin}
\begin{tabular}{cccc}
\hline
\multicolumn{1}{c}{parameter} &\multicolumn{1}{c}{$\sigma_{11}$}
&\multicolumn{1}{c}{$\sigma_{10}$}&\multicolumn{1}{c}{$\sigma_{01}$}\\
\hline
\multicolumn{1}{l}{$\alpha$} &\multicolumn{1}{c}{3.772}
&\multicolumn{1}{c}{15.28}&\multicolumn{1}{c}{146.3}\\
\multicolumn{1}{l}{$\beta$} &\multicolumn{1}{c}{1.262}
&\multicolumn{1}{c}{0}&\multicolumn{1}{c}{0}\\
\multicolumn{1}{l}{$m_{0}$(MeV)}
 &\multicolumn{1}{c}{1188}
&\multicolumn{1}{c}{1245}&\multicolumn{1}{c}{1472}\\
\multicolumn{1}{l}{$\Gamma$(MeV)} &\multicolumn{1}{c}{99.02}
&\multicolumn{1}{c}{137.4}&\multicolumn{1}{c}{26.49}\\
\hline
\end{tabular}
\end{table}

\begin{table}
\caption{Higher $N^{*}$ resonances}
\label{nres}
\begin{tabular}{lcccccr}
\hline
R & $J^{p}$ &Mass &Width &$N\pi$ &$\Delta\pi$ &$N\rho$\\
\hline
$D_{13}$ &$\frac{3}{2}^{-}$&1.520 &0.125
& 0.550&0.250&0.200
\\
\multicolumn{1}{l}{$S_{11}$} &\multicolumn{1}{c}{$\frac{1}{2}^{-}$}
&\multicolumn{1}{c}{1.650}&\multicolumn{1}{c}{0.150}
&\multicolumn{1}{c}{0.600}&\multicolumn{1}{c}{0.050}
&\multicolumn{1}{c}{0.180}
\\
\multicolumn{1}{l}{$D_{15}$} &\multicolumn{1}{c}{$\frac{5}{2}^{-}$}
&\multicolumn{1}{c}{1.675}&\multicolumn{1}{c}{0.155}
&\multicolumn{1}{c}{0.380}&\multicolumn{1}{c}{0.580}
&\multicolumn{1}{c}{0.030}
\\
\multicolumn{1}{l}{$F_{15}$} &\multicolumn{1}{c}{$\frac{5}{2}^{+}$}
&\multicolumn{1}{c}{1.680}&\multicolumn{1}{c}{0.125}
&\multicolumn{1}{c}{0.600}&\multicolumn{1}{c}{0.125}
&\multicolumn{1}{c}{0.125}
\\
\multicolumn{1}{l}{$D_{13}$} &\multicolumn{1}{c}{$\frac{3}{2}^{-}$}
&\multicolumn{1}{c}{1.700}&\multicolumn{1}{c}{0.100}
&\multicolumn{1}{c}{0.100}&\multicolumn{1}{c}{0.380}
&\multicolumn{1}{c}{0.100}
\\
\multicolumn{1}{l}{$P_{11}$} &\multicolumn{1}{c}{$\frac{1}{2}^{+}$}
&\multicolumn{1}{c}{1.710}&\multicolumn{1}{c}{0.110}
&\multicolumn{1}{c}{0.150}&\multicolumn{1}{c}{0.100}
&\multicolumn{1}{c}{0.200}
\\
\multicolumn{1}{l}{$P_{13}$} &\multicolumn{1}{c}{$\frac{3}{2}^{+}$}
&\multicolumn{1}{c}{1.720}&\multicolumn{1}{c}{0.200}
&\multicolumn{1}{c}{0.150}&\multicolumn{1}{c}{0.100}
&\multicolumn{1}{c}{0.530}
\\
\multicolumn{1}{l}{$F_{17}$} &\multicolumn{1}{c}{$\frac{7}{2}^{+}$}
&\multicolumn{1}{c}{1.990}&\multicolumn{1}{c}{0.290}
&\multicolumn{1}{c}{0.050}&\multicolumn{1}{c}{0.060}
&\multicolumn{1}{c}{0.340}
\\
\hline
\end{tabular}
\end{table}

\begin{table}
\caption{Higher $\Delta$ resonances}
\label{dres}
\begin{tabular}{lcccccr}
\hline
R & $J^{p}$ &Mass &Width &$N\pi$ &$\Delta\pi$ &$N\rho$\\
\hline
$P_{33}$ &$\frac{3}{2}^{+}$&1.600 &0.250
& 0.350&0.450&0.050
\\
\multicolumn{1}{l}{$S_{31}$} &\multicolumn{1}{c}{$\frac{1}{2}^{-}$}
&\multicolumn{1}{c}{1.620}&\multicolumn{1}{c}{0.160}
&\multicolumn{1}{c}{0.300}&\multicolumn{1}{c}{0.600}
&\multicolumn{1}{c}{0.070}
\\
\multicolumn{1}{l}{$D_{33}$} &\multicolumn{1}{c}{$\frac{3}{2}^{-}$}
&\multicolumn{1}{c}{1.700}&\multicolumn{1}{c}{0.280}
&\multicolumn{1}{c}{0.150}&\multicolumn{1}{c}{0.698}
&\multicolumn{1}{c}{0.150}
\\
\multicolumn{1}{l}{$S_{31}$} &\multicolumn{1}{c}{$\frac{1}{2}^{-}$}
&\multicolumn{1}{c}{1.900}&\multicolumn{1}{c}{0.150}
&\multicolumn{1}{c}{0.100}&\multicolumn{1}{c}{0.050}
&\multicolumn{1}{c}{0.450}
\\
\multicolumn{1}{l}{$F_{35}$} &\multicolumn{1}{c}{$\frac{5}{2}^{+}$}
&\multicolumn{1}{c}{1.905}&\multicolumn{1}{c}{0.300}
&\multicolumn{1}{c}{0.100}&\multicolumn{1}{c}{0.250}
&\multicolumn{1}{c}{0.450}
\\
\multicolumn{1}{l}{$P_{31}$} &\multicolumn{1}{c}{$\frac{1}{2}^{+}$}
&\multicolumn{1}{c}{1.910}&\multicolumn{1}{c}{0.220}
&\multicolumn{1}{c}{0.220}&\multicolumn{1}{c}{0.089}
&\multicolumn{1}{c}{0.060}
\\
\multicolumn{1}{l}{$P_{33}$} &\multicolumn{1}{c}{$\frac{3}{2}^{+}$}
&\multicolumn{1}{c}{1.920}&\multicolumn{1}{c}{0.250}
&\multicolumn{1}{c}{0.200}&\multicolumn{1}{c}{0.190}
&\multicolumn{1}{c}{0.080}
\\
\multicolumn{1}{l}{$D_{35}$} &\multicolumn{1}{c}{$\frac{5}{2}^{-}$}
&\multicolumn{1}{c}{1.930}&\multicolumn{1}{c}{0.250}
&\multicolumn{1}{c}{0.090}&\multicolumn{1}{c}{0.200}
&\multicolumn{1}{c}{0.120}
\\
\multicolumn{1}{l}{$F_{37}$} &\multicolumn{1}{c}{$\frac{7}{2}^{+}$}
&\multicolumn{1}{c}{1.950}&\multicolumn{1}{c}{0.240}
&\multicolumn{1}{c}{0.400}&\multicolumn{1}{c}{0.130}
&\multicolumn{1}{c}{0.080}
\\
\hline
\end{tabular}
\end{table}

\section*{Figure Captions}
\begin{description}

\item{\bf Fig.\ 1}\ \ \
$pp$ inelastic cross sections as functions of
center-of-mass energy $\sqrt{s}$.
\item{\bf Fig.\ 2}\ \ \
$pp\rightarrow pp\omega$ and $pp\rightarrow pp\rho^0$ cross sections as
functions of center-of-mass energy.

\item{\bf Fig.\ 3}\ \ \
The $\pi^-+p$ elastic cross section as a function of center-of-mass
energy. The solid line is the experimental
data and the dotted line is the contribution from the formation of
$\Delta, N^*(1440)$ and $N^*(1535)$ resonances. The dashed line is the
difference between the solid and dotted lines and is attributed to
direct $\pi+N$ scattering.

\item{\bf Fig.\ 4}\ \ \
$\pi^++p$ elastic cross section as a function of center-of-mass
energy. The solid line is the experimental
data and the dotted line is the contribution from the formation of
the $\Delta$ resonance. The dashed line is the
difference between the solid and dotted lines and is attributed to
direct $\pi+N$ scattering.

\item{\bf Fig.\ 5}\ \ \
The $\pi^0+p(n)$ elastic cross section
as a function of center-of-mass energy calculated from the resonance model.

\item{\bf Fig.\ 6}\ \ \
The $\pi^++\Delta^-(\Delta^0)$ elastic cross section
as a function of center-of-mass energy calculated from the resonance model.

\item{\bf Fig.\ 7}\ \ \
The $\pi^++p$ inelastic cross section as a function of
the center-of-mass energy $\sqrt{s}$.

\item{\bf Fig.\ 8}\ \ \
The isospin-averaged $\pi\pi$ elastic (upper window)
and inelastic (lower window) cross sections as functions of
center-of-mass energy $\sqrt{s}$.

\item{\bf Fig.\ 9}\ \ \
The angular distribution of $NN\rightarrow N\Delta$ reaction at
different center-of-mass energies.

\item{\bf Fig.\ 10}\ \ \
Transverse momentum distributions for protons (upper window) and
pions (lower window) from $pp$ collisions at $p_{beam}$=15 GeV/c
and 3 GeV/c. Solid lines are the scaling results valid at high energies
as discussed in the text.

\item{\bf Fig.\ 11}\ \ \
The average energy per nucleon (upper window) and
the adiabatic sound velocity (lower window)
as functions of density in nuclear matter at zero temperature.

\item{\bf Fig.\ 12}\ \ \
Reaction rates for various channels in nucleon+nucleon collisions
as functions of time in head-on collisions of Au+Au at
$p_{beam}/A=$ 11.6 GeV/c.

\item{\bf Fig.\ 13}\ \ \
Reaction rates of nucleon+baryon resonance, baryon resonance+baryon
resonance and meson+meson collisions
as functions of time in head-on collisions of Au+Au at
$p_{beam}/A=$ 11.6 GeV/c.

\item{\bf Fig.\ 14}\ \ \
Rates of baryon resonance decays and formations in meson+baryon
collisions as functions of time in head-on collisions of Au+Au at
$p_{beam}/A=$ 11.6 GeV/c.

\item{\bf Fig.\ 15}\ \ \
The evolution of the $\pi, \Delta$, and $N^*$ multiplicities
in head-on collisions of Au+Au at $p_{beam}/A=$ 11.6 GeV/c.

\item{\bf Fig.\ 16}\ \ \
Collision number distributions as functions of center-of-mass energy
in head-on collisions of Au+Au at
$p_{beam}/A=$ 11.6 GeV/c.

\item{\bf Fig.\ 17}\ \ \
The evolution of local baryon density in the reaction plane
in head-on collisions of Au+Au at
$p_{beam}/A=$ 11.6 GeV/c.

\item{\bf Fig.\ 18}\ \ \
Same as Fig. 17 for the scalar baryon density.

\item{\bf Fig.\ 19}\ \ \
Same as Fig. 17 for the local meson density.

\item{\bf Fig.\ 20}\ \ \
Same as Fig. 17 for the local energy density.

\item{\bf Fig.\ 21}\ \ \
The evolution of central local baryon and
energy densities in the collision of
Au+Au at $P_{beam}/A=$11.6 GeV/c and b=0 calculated with the cascade,
the soft, and the stiff equation of state.

\item{\bf Fig.\ 22}\ \ \
The evolution of the fraction of particles
with local energy density higher than 2.5 GeV/fm$^3$
in head-on collisions of Au+Au at $P_{beam}/A=$11.6 GeV/c.

\item{\bf Fig.\ 23}\ \ \
(Upper window): the evolution of the particle number
in a sphere with a radius of 2 fm around the center of mass.
(Middle window): the evolution of the average kinetic energy
$< \frac{2}{3}E_K >$
per particle in the sphere. (Lower window): the evolution of the
stopping ratio R in head-on collisions of
Au+Au at $P_{beam}/A=$11.6 GeV/c.

\item{\bf Fig.\ 24}\ \ \
(Upper windows): Radial density distributions for baryons and pions at
freeze-out calculated using the cascade (left) and the soft equation of
state (right). (Lower windows): the radial flow velocity profiles at
freeze-out in head-on collisions of Au+Au at $P_{beam}/A=$11.6 GeV/c.

\item{\bf Fig.\ 25}\ \ \
Transverse momentum distributions of protons
in the collision of Au+Au at $P_{beam}/A=$11.6 GeV/c and
at an impact parameter of 2 fm.
The solid line is the least square fit to preliminary data
from the E802 collaboration. See discussions in the text.

\item{\bf Fig.\ 26}\ \ \
Rapidity distributions of protons
in the collision of Au+Au at $P_{beam}/A=$11.6 GeV/c and
at an impact parameter of 2 fm.

\item{\bf Fig.\ 27}\ \ \
Comparison of rapidity distributions of protons
in the collision of Au+Au at $P_{beam}/A=$11.6 GeV/c and at
impact parameters of 1 and 3 fm.

\item{\bf Fig.\ 28}\ \ \
Transverse momentum distributions of pions from the cascade model
for Au+Au collisions at $P_{beam}/A=$11.6 GeV/c and at
an impact parameter of 2 fm.
Solid lines are least square fits to preliminary
data from the E802/E866 collaboration.

\item{\bf Fig.\ 29}\ \ \
Same as Fig. 28 for transverse momentum distributions of pions.

\item{\bf Fig.\ 30}\ \ \
Rapidity distributions of $\pi^+$
in the collision of Au+Au at $P_{beam}/A=$11.6 GeV/c at
an impact parameter of 2 fm.

\item{\bf Fig.\ 31}\ \ \
Comparison of rapidity distributions of $\pi^+$
in the collision of Au+Au at $P_{beam}/A=$11.6 GeV/c and at
impact parameters of 1 and 3 fm.

\item{\bf Fig.\ 32}\ \ \
Energy distributions of $K^+$ producing collisions
as functions of center-of-mass energy
in head-on collisions of Au+Au at
$p_{beam}/A=$ 11.6 GeV/c.

\item{\bf Fig.\ 33}\ \ \
The evolution of $K^+$ production rates
in head-on collisions of Au+Au at $p_{beam}/A=$ 11.6 GeV/c.

\item{\bf Fig.\ 34}\ \ \
Transverse momentum distributions of $K^+$
in the collision of Au+Au at $P_{beam}/A=$11.6 GeV/c
at an impact parameter of 2 fm.
Solid lines are least square fits to preliminary
data (solid symbols) from the E802/E866 collaboration.

\item{\bf Fig.\ 35}\ \ \
Rapidity distributions of $K^+$
in the collision of Au+Au at $P_{beam}/A=$11.6 GeV/c
at an impact parameter of 2 fm.

\item{\bf Fig.\ 36}\ \ \
Comparison of rapidity distributions of $K^+$
in the collision of Au+Au at $P_{beam}/A=$11.6 GeV/c
and at impact parameters of 1 and 3 fm.

\item{\bf Fig.\ 37}\ \ \
Baryon average transverse velocity in the reaction plane as a function of
rapidity for Au+Au collisions at $P_{beam}/A$=11.6 GeV/c and
at impact parameters of 2, 6, and 10 fm.

\item{\bf Fig.\ 38}\ \ \
Time evolution of the flow parameter, the central density and the total
in-plane
transverse momentum in the collision of
Au+Au at $P_{beam}/A$=11.6 GeV/c and at an impact parameter
of 2 fm.

\item{\bf Fig.\ 39}\ \ \
Same as Fig. 38 at an impact parameter of 6 fm.

\end{description}


\begin{thebibliography}{99}

\bibitem{qmatter} Proceedings of Heavy Ion Physics at the AGS,
HIPAGS'93, 13-15, Jan., 1993,
Eds.\ G.S.F. Stephans, S.G. Steadman, and W.L. Kehoe;
Quark Matter 93, Nucl. Phys. {\bf A566}, 1c (1994).

\bibitem{rqmd}H. Sorge, H. St\"ocker, and W. Greiner, Ann. of Phys. (NY)
{\bf 192}, 266 (1989); R. Mattiello, H. Sorge, H. St\"ocker, and W. Greiner,
Phys.\ ReV.\ Lett.\ {\bf 63}, 1459 (1989); H. Sorge, A.V. Keitz, L.
Winckelmann,
A. Jahns, H. Sorge, H. St\"ocker, and W. Greiner,
Phys.\ Lett.\ {\bf B263}, 353 (1991); M. Hofmann, R. Mattiello, H. Sorge,
H. St\"ocker and W. Greiner Phys.\ Rev.\ C{\bf 51}, 2095 (1995) and
references therein.

\bibitem{arc} Y. Pang, T.J. Schlagel, and S.H. Kahana,
Phys.\ Rev.\ Lett.\ {\bf 68}, 2743 (1992); T.J. Schlagel, Y. Pang
and S.H. Kahana, Phys.\ Rev.\ Lett.\ {\bf 69}, 3290 (1992);
S.H. Kahana, Y. Pang, T. J. Schlagel and C. Dover,
Phys.\ Rev.\ C {\bf 47}, R1356 (1993).

\bibitem{qgsm}L. Bravina, L.P. Csernai, P. Levai and D. Strottman,
Phys.\ Rev.\ C {\bf 50}, 2161 (1994); and references therein.

\bibitem{kapusta}J.I. Kapusta, A. P. Vischer and R. Venugopalan,
Phys.\ Rev.\ C {\bf 51}, 901 (1995).

\bibitem{e866n}F. Videbak et al.\, for the E802/E866 collaboration,
	in: Proc. Quark Matter '95, Monterey, CA, Jan. 9-13, 1995,
	Nucl.\ Phys.\ {\bf A} in press.

\bibitem{e877n}J. Barrette et al.\, for the E877 collaboration,
	in: Proc. Quark Matter '95, Monterey, CA, Jan. 9-13, 1995,
	Nucl.\ Phys.\ {\bf A} in press.

\bibitem{eosn}P. Bradly et al.\ for the EOS collaboration,
	AGS proposal ``Exclusive Study of Nuclear Collisions at the AGS'',
	LBL preprint (1993).

\bibitem{li95}B.A. Li, C.M. Ko and G.Q. Li, submitted to Phys.\ Rev.\ Lett.\

\bibitem{keywest}B.A. Li, C. M. Ko and G.Q. Li, in: Proceedings of
the 11th Winter Workshop on Nuclear Dynamics, Feb. 11-18, 1995,
Key West, Florida, Eds.\ A. Mignery et al.

\bibitem{bertsch}G.F. Bertsch and S. Das Gupta,
Phys. Rep., {\bf 160}, 189 (1988).

\bibitem{li91a}B.A. Li and W. Bauer, Phys.\ Lett.\ {\bf B254}, 335 (1991);
        Phys. Rev.\ C {\bf 44}, 450 (1991).
\bibitem{li91b} B.A. Li, W. Bauer and G.F. Bertsch,
Phys.\ Rev.\ C {\bf 44}, 2095 (1991).

\bibitem{data1}Total cross-sections for the reactions of high energy
particles, A. Baldini, V. Flaminio, W.G. Moorhead, D.R.O. Morrison,
(Springer-Verlag, Berlin), 1988.

\bibitem{verwest}B.J. VerWest and R.A. Arndt, Phys. Rev.
C {\bf 25}, 1979 (1980).

\bibitem{wolf90}Gy. Wolf, G. Batko, W. Cassing and U. Mosel,
        Nucl.\ Phys.\ {\bf A517}, 615 (1990).

\bibitem{dani91}P. Danielewicz and G.F. Bertsch,
        Nucl.\ Phys.\ {\bf A533}, 712 (1991).

\bibitem{wolf}Gy. Wolf, W. Cassing and U. Mosel,
        Nucl.\ Phys.\ {\bf A552}, 549 (1993).

\bibitem{aichelin}S. Huber and J. Aichelin,
Nucl.\ Phys.\ {\bf A573}, 587 (1994).


\bibitem{li93}B.A. Li, Nucl.\ Phys.\ {\bf A552}, 605 (1993).

\bibitem{pilkuhn}H.M. Pilkuhn, Relativistic particle physics,
(Spring-Verlag) 1979.


\bibitem{bert}G.F. Bertsch, M. Gong, L. McLerran, V. Ruuskanen and E.
Sarkkinen,
Phys.\ Rev.\ D {\bf 37}, 1202 (1988).

\bibitem{ko}C.M. Ko, Z.G. Wu, L.H. Xia and G.E. Brown, Phys.\ Rev.\ Lett.\
{\bf 66}, 2577 (1991); Phys.\ Rev.\ C {\bf 43}, 1881 (1991).

\bibitem{lbl} O. Benary, R. Price and G. Alexander, NN and ND
interactions (above 0.5 GeV/c) -a compilation, Berkeley (1970).

\bibitem{rand}J. Randrup and C.M. Ko, Nucl.\ Phys.\ {\bf A343}, 519 (1980).

\bibitem{fasselor} K. Tsushima, S.W. Huang and A. Faessler,
Phys.\ Lett.\ {\bf B337}, 245 (1994).

\bibitem{cugnon}J. Cugnon and R.M. Lombard,
        Nucl.\ Phys.\ {\bf A422}, 635 (1984).

\bibitem{liko}G. Q. Li and C. M. Ko, Nucl. Phys. {\bf A}, submitted.

\bibitem{data2}Properties and production spectra of elementary particles,
A. N. Diddens, H. Pilkuhn and K. Schl\"upmann,
(Springer-Verlag, Berlin), 1972.

\bibitem{wong}C.Y. Wong, Introduction to High Energy Heavy-Ion Collisions,
(World Scientific, Singapore), 1994.

\bibitem{kapusta1} C. Grant and J. Kapusta, Phys.\ Rev.\ C{\bf 32}, 663 (1985).
\bibitem{strotman}E. Osnes and D. Strottman, Phys.\ Lett.\ {\bf 166B}, 5
(1986).
\bibitem{ginocchio} J.N. Ginocchio, Phys.\ Rev.\ C{\bf 17}, 195 (1978).
\bibitem{baldo}M. Baldo and L.S. Ferreira,
                Nucl.\ Phys. {\bf A569}, 645 (1994)
\bibitem{betz}M. Betz and T.-S.H. Lee,
                Phys.\ Rev.\ C{\bf 23}, 375 (1981).
\bibitem{lee}T.S.-H. Lee and K. Ohta,
                Phys.\ Rev.\ C{\bf 25}, 3043 (1982).
\bibitem{haar}B.ter Haar and R. Malfliet,
                Phys.\ Rep.\ {\bf 149}, 287 (1987).
\bibitem{esbensen}H. Esbensen and T.-S.H. Lee,
                Phys.\ Rev.\ C {\bf 32}, 1966 (1985).
\bibitem{connel}J.S. O'Connell and R.M. Sealock,
                Phys.\ Rev.\ C{\bf 42}, 2290 (1990).
\bibitem{horikawa}Y. Horikawa, M. Thies and F. Lenz,
                Nucl.\ Phys.\ {\bf A345}, 386 (1980).
\bibitem{johnson}M.B. Johnson and D.J. Ernst,
                Ann.\ Phys.\ (N.Y) 219, 266 (1992).
\bibitem{li93a}B.A. Li, Phys.\ Lett.\ {\bf B292}, 246;
        {\bf B300}, 14 (1993); Phys.\ Rev.\ C{\bf 47}, 693 (1993).
\bibitem{li93b}B.A. Li, Phys.\ Lett.\ {\bf B319}, 412 (1993);
        Nucl.\ Phys.\ {\bf A570}, 797 (1994).
\bibitem{brown}G.E. Brown, Nucl. Phys. {\bf A 522}, 397c (1991);\\
G.E. Brown and M. Rho, Phys. Rev. Lett. {\bf 66}, 2720 (1991).

\bibitem{kaplan}D.B. Kaplan and A.E. Nelson, Phys. Lett. B175 (1986) 57;\\
A.E. Nelson and D.B. Kaplan, Phys. Lett. {\bf B 192}, 193 (1987).

\bibitem{hatsuda}T. Hatsuda and S.H. Lee, Phys. Rev. C{\bf 46}, R34 (1992).

\bibitem{asakawa} M. Asakawa, C.M. Ko, P. L\'evai and X.J. Qiu,
        Phys. Rev. C {\bf 46}, R1159 (1992);\\
M. Asakawa and C.M. Ko, Nucl. Phys. A560, 399 (1993).

\bibitem{dani95}P. Danielewicz, Phys.\ Rev.\ C{\bf 51}, 716 (1995).

\bibitem{li94}B.A. Li, C.M. Ko and G.Q. Li,
Phys.\ Rev.\ C{\bf 50}, R2675 (1994).

\bibitem{rischke}D.H. Rischke, H. St\"ocker, and W. Greiner,
J. Phys. G14, 191, (1988); Phys.\ Rev.\ D{\bf 41}, 111 (1990).

\bibitem{glen}N.K. Glendenning, Nucl. Phys.\ {\bf A512}, 737 (1990).

\bibitem{gyu}M. Gyulassy, Quark Matter 95: Concluding remarks.

\bibitem{wong82} C.Y. Wong, Phys.\ Rev.\ C{\bf 25}, 1461 (1982).

\bibitem{munz} P. Braun-Munzinger, J. Stachel, J.P. Wessels and N. Xu,
Phys.\ Lett. {\bf B333}, 33 (1995).

\bibitem{e866} Michel Gonin for the E802/E866 collaboration,
in: Proceedings of Heavy-ion Physics at the AGS, 13-15 Jan. 1993,
Eds.\ G. S.F. Stephans, S.G. Steadman and W.L. Kehoe;
Nucl. Phys.\ {\bf A553}, 799c (1993).

\bibitem{raf} J. Rafelski and B. M\"uller, Phys.\ Rev.\ Lett.\
{\bf 48}, 1066 (1982).

\bibitem{koxia} C.M. Ko and L.H. Xia, Phys.\ Lett.\ {\bf B222}, 343 (1989 ).

\bibitem{aich} J. Aichelin and C. M. Ko, Phys.\ Rev.\ Lett.\
{\bf 55}, 2661 (1985).

\bibitem{huang} S.W. Huang et al.\, Phys.\ Lett.\ {\bf B298}, 41 (1993).

\bibitem{fang} X.S. Fang, C. M. Ko, G.Q. Li and Y.M. Zhang,
Phys.\ Rev.\ C{\bf 49}, R608 (1994); Nucl.\ Phys.\ {\bf A575}, 766 (1994).

\bibitem{mar}T. Maruyama, W. Cassing, U. Mosel, S. Teis and K. Weba,
Nucl.\ Phys.\ {\bf A573}, 653 (1994).

\bibitem{likaon}B.A. Li, Phys.\ Rev.\ C{\bf 50}, 2144 (1994).

\bibitem{dani} P. Danielewicz and G. Odyniec,
Phys.\ Lett.\ {\bf B157}, 146 (1985).

\bibitem{gale}J. Zhang, S. Das Gupta and C. Gale,
Phys.\ Rev.\ C{\bf 50}, 1617 (1994).

\bibitem{E877}J. Barrette et. al. for the E877 collaboration,
Phys.\ Rev.\ Lett.\ {\bf 73}, 2532 (1994).

\bibitem{E877-95}J. P. Wessels, in: Proceedings of
the 11th Winter Workshop on Nuclear Dynamics, Feb. 11-18, 1995,
Key West, Florida, Eds.\ A. Mignery et al.

\bibitem{zhang}Y. Zhang et al.\, for the E877 collaboration,
	in: Proc. Quark Matter '95, Monterey, CA, Jan. 9-13, 1995,
	Nucl.\ Phys.\ {\bf A} in press.

\bibitem{xiong}L. Xiong, C.M. Ko and V. Koch,
Phys.\ Rev.\ C{\bf 47}, 788 (1993).

\bibitem{koch}V. Koch and G.F. Bertsch, Nucl.\ Phys.\ {\bf A552}, 591 (1993).
\bibitem{gqli}G. Q. Li, C. M. Ko and B.A. Li,
Phys.\ Rev.\ Lett.\ {\bf 74}, 235 (1995).

\bibitem{gerd}G. Welke and G.F. Bertsch, Phys.\ Rev.\ C{\bf 45}, 1403 (1992);
G.F. Bertsch, in Statistical Description of Transport in Plasma, Astro-and
Nuclear Physics, Eds.\ J. Misquich, G. Pelletier and P. Schuck (Nova Science
Publishers, INC) P. 337 (1993).
\end{thebibliography}
\end{document}